\documentstyle[epsfig,12pt,preprint,tighten,aps]{revtex}
\begin{document}

\title{
\rightline{{\tt March 2001}}
\ \\
Active-sterile neutrino oscillations in the early Universe: asymmetry generation at
low $|\delta {\rm m}^2|$ and the Landau-Zener approximation}
\author{P. Di Bari$^{1,2}$ and R. Foot$^2$}
\maketitle
\begin{center}
{\em
$^1$ Istituto Nazionale di Fisica Nucleare (INFN), Italy\\
$^2$ School of Physics \\
Research Centre for High Energy Physics\\
The University of Melbourne\\
Victoria 3010 Australia\\
(dibari,foot@physics.unimelb.edu.au)}
\end{center}

\begin{abstract} 
It is well established that active-sterile 
neutrino oscillations generate
large neutrino asymmetries for very small mixing 
angles ($\sin^2 2\theta_0\lesssim 10^{-4}$),
negative values of $\delta m^2$ and provided 
that $|\delta m^2|\gtrsim 10^{-4}\,{\rm eV^2}$. 
By numerically solving the quantum kinetic equations,
we show that the generation still occurs at much lower 
values of $|\delta m^2|$. We also describe the borders of the 
generation at small mixing angles and
show how our numerical results can be analytically understood
within the framework of the Landau-Zener approximation
thereby extending previous work based on the adiabatic limit.
This approximate approach leads
to a fair description of the MSW dominated regime of the neutrino 
asymmetry evolution and is also
able to correctly reproduce its final value.
We also briefly discuss the impact that neutrino asymmetry generation 
could have on big bang nucleosynthesis, CMBR and relic neutrinos.
\end{abstract}
\newpage

\section{Introduction}

Previous work has established 
that $\nu_{\alpha}-\nu_{\rm sterile}\,(\alpha=e,\mu,\tau)$ 
neutrino oscillations typically generate large [${\cal O}(0.1)$]
neutrino asymmetries for {\em negative values} of $\delta m^2$
\footnote{We define $\delta m^2\equiv m^2_2-m^2_1$, with $m_1$ and $m_2$
eigenvalues of the mass eigenstates such that in the limit of
$\theta_0\rightarrow 0$ they coincide with the $\alpha$ and sterile weak eigenstates
respectively. 
With this definition in mind, one can say that {\em negative values
of $\delta m^2$ imply a sterile neutrino lighter than the $\alpha$.}}
and {\em small mixing angles} ($\sin^2 2\theta_0\ll 
1$) \cite{ftv,fv1,fv2,f98}.
Such a large value for the neutrino asymmetry can have several
interesting consequences and applications including: a suppression of the 
sterile neutrino production
\cite{fv,fv1,dll}, a modification of the standard BBN predictions in different ways
\cite{fv1,shi,fv2,bfv,foot,domains} 
\footnote{A particular interesting case is the possibility that a
neutrino asymmetry generation can solve the  
claimed discrepancy between the value of baryon density 
from standard BBN with that one inferred from  recent CMBR anisotropies 
observations \cite{ropa2}.} 
and, perhaps not emphasized enough, it implies a relic 
$\alpha$ neutrino background in which mainly neutrinos or anti-neutrinos 
(according to the sign of the generated asymmetry) are present.

We define the {\em asymmetry} of the particle species $X$ as 
{\em the net number of $X$ particles at temperature $T$ 
per number of photons at some fixed initial temperature $T_{\rm in}$,
such that $m_{\mu} \gg T_{\rm in} \gg m_{\rm el}$, in a comoving volume $R^3$}:
\begin{equation}\label{LX}
L_X\equiv {N_X-N_{\bar{X}}\over N_{\gamma}^{\rm in}},
\end{equation} 
where $N_X\equiv n_X\,R^3$ is the number of $X$-particles in $R^3$, while $n_x$ 
is  the $X$-particle density
\footnote{If one defines:
\begin{equation}
T_{\nu}\equiv T_{\rm in}\,{R_{\rm in}\over R}
\end{equation}
it is simple to chech that the asymmetry $L_{X}$ is connected to the 
{\em asymmetry abundance} (relative to photons) 
$\eta_X\equiv (n_X-n_{\bar{X}})/n_{\gamma}$ by 
the simple relation:
\begin{equation}
\eta_X=h\,L_X  ,
\end{equation}
where $h\equiv (T_{\nu}/T)^3 \leq 1$ is the {\em dilution factor} 
that takes into account the photon production
during the electron-positron annihilations around $m_e/2\simeq 0.25{\rm MeV}$.  
}. 

The exact {\em borders} of the generation in the space of mixing parameters
have not yet been fully determined. 
Using the so called ``static approximation'' in \cite{fv1}  
it was found that, to have a final neutrino asymmetry larger than $10^{-5}$,
the following rough {\em lower limit on the mixing angle} 
(for a fixed $\delta m^2$) holds:
\begin{equation}\label{lower}
\sin^2{2\theta_0}\gtrsim 6(5)\times 10^{-10}\,
\left(|\delta m^2|\over{\rm eV}^2\right)^{-{1\over 6}}
\hspace{10mm} \alpha=e\,(\mu,\tau)
\end{equation}
An {\em upper limit on the mixing angle} for which the neutrino generation occurs
is still unknown as numerical solutions are made quite difficult
(i.e. CPU time consuming) by the presence
of a sterile neutrino production prior the onset of neutrino asymmetry generation
that is strongly coupled to the neutrino asymmetry generation. Moreover,
at large angles ($\sin^2 2\theta_0\gtrsim 10^{-6}$), the phenomenon of rapid oscillations of the asymmetry at the onset of the generation appears \cite{ropa}, 
making things even more complicated 
\footnote{Actually the appearance of neutrino asymmetry sign oscillations, 
in the region where the sterile neutrino production is not negligible, may 
not be a coincidence. It is likely that the strong coupling between a 
copious sterile neutrino production and the neutrino asymmetry evolution 
is responsible for the rapid sign changes \cite{ben}.}.
Requiring that the final effective number of extra-neutrino species (in energy density) 
does not violate the BBN bound, $\Delta N^{\rho}_{\nu}< \Delta N^{\rm max}_{\nu}<1$, 
and with the definition 
$f(\Delta N^{\rm max}_{\nu})\equiv \ln(1-\Delta N^{\rm max}_{\nu})/\ln(0.4)$,
one finds, for $\alpha=e\,(\mu,\tau)$, the constraint \cite{fv1,dll} :
\begin{equation}\label{upper}
\sin^2 2\theta_0 \leq 2.1\,(4.1)\times 10^{-5}\,
\left({\rm eV}^2\over |\delta m^2| \right)^{1\over 2}
\,f^2(\Delta N^{\rm max}_{\nu}) , 
\end{equation}
valid for  $|\delta m^2|\gtrsim 10^{-4}\,{\rm eV}^2$ (with $\delta m^2<0$)  and 
$0.9\gtrsim \Delta N^{\rm max}_{\nu}\gtrsim 0.3$. 
This constraint is however applicable only in the case of two flavour 
$\nu_{\alpha}-\nu_s$ (with $\alpha=\mu,\tau$)
neutrino oscillations. In the general case of multiflavor oscillations
the neutrino asymmetry produced by the oscillations with the largest 
$|\delta m^2|$ can suppress the sterile neutrino production from the other
oscillations \cite{fv,fv1,dll}. Therefore Eq. (\ref{upper}) can only 
possibly apply for oscillations with the largest $|\delta m^2|$, with the
constraints on other two flavor oscillations being much weaker. 
Also, if $\alpha=e$, or if the muon or tauon neutrino is also mixed to the 
electron neutrino then an electron neutrino asymmetry can be generated, the constraint 
$\Delta N^{\rho}_{\nu}< \Delta N_{\nu}^{\rm max}$ has to be replaced by a more general 
constraint $\Delta N_{\nu}^{\rho}+\Delta N_{\nu}^{f_{\nu_e}}< \Delta N_{\nu}^{\rm max}$, 
in which the quantity $\Delta N^{f_{\nu_e}}_{\nu}$ 
takes into account the BBN effect of the 
deviation of the electron neutrino distribution from the 
standard case, in which a Fermi-Dirac distribution with zero chemical potential is assumed
\footnote{A more detailed discussion on the definition of $\Delta N_{\nu}^{\rho}$ and
$\Delta N_{\nu}^{f_{\nu_e}}$ can be found in \cite{ropa2}.}. 
If a positive electron neutrino asymmetry is generated before the
freezing of the neutron to proton abundance ratio,
then $\Delta N_{\nu}^{f_{\nu_e}}$ is negative \cite{ftv,shi,fv2,foot,ropa2}. 
In this case the upper limit
on the mixing angle can be strongly relaxed and is related to the limit   
for the neutrino asymmetry generation, that however, as we said, has not been
determined. Moreover one has conservatively to keep in mind 
that the current observational situation on the values of primordial nuclear
abundances cannot strictly exclude a value $\Delta N_{\nu}^{\rm max}\geq 1$, for which
no BBN constraints on active-sterile neutrino oscillations would apply at all if
$\Delta N^{f_{\nu_e}}_{\nu}=0$
\footnote{More generally no BBN constraint applies if 
$\Delta N_{\nu}^{\rm max}\geq 1+{\rm max}(\Delta N_{\nu}^{f_{\nu_e}})$.}.  
It has also to be mentioned that the region of the mixing parameter space 
in which a positive electron neutrino asymmetry is generated such that 
$\Delta N_{\nu}^{f_{\nu_e}}$ is negative
and with a mixing angle large enough that $\Delta N^{\rho}_{\nu}\sim 1$, would be very interesting as  in this case the standard BBN would be modified 
such that the
primordial nuclear abundances predictions would still agree with the measured values  
and at the same time a higher value for the baryon density would be allowed,
in agreement with that one inferred from CMBR acoustic peaks 
\cite{BoomMax,ropa2}. 

For 
\begin{equation}\label{lowerde} 
|\delta m^2| \gtrsim 10^{-4}{\rm eV}^2,
\end{equation} 
the numerical calculations performed with the quantum kinetic equations (QKE)  
\cite{QKE} found that the final value
of the neutrino asymmetry is in the range $0.23-0.35$ approximately
independently
of $\sin 2\theta_0$ for most of the range of values
set by Eqs.(\ref{lower},\ref{upper})\cite{fv2}.
These calculations also showed that the final values can be well reproduced
within the {\em adiabatic approximation}. Let us shortly recall 
the main features
of the neutrino asymmetry generation effect and of what the 
adiabatic approximation consists
\footnote{See also \cite{comment} for a more detailed review.}. 
Neutrino asymmetry evolution in the region of mixing parameters
set by the equations (\ref{lower}), (\ref{upper}) and (\ref{lowerde}), 
is characterized by the presence of a {\em critical temperature} $T_c$ that is 
approximately given by the following expression:
\begin{equation}\label{Tc}
T_c \simeq T_{\rm res}^0(y_c)\equiv 15.0\,(18.6)\,
\left({|\delta m^2| \over \text{eV}^2}\right)^{1 \over 6}
\,\left({2\over y_c}\right)^{1\over 3} \text{MeV},
\end{equation}
where $y_c \equiv p_c/T$ is 
the {\em critical (dimensionless) resonant momentum} and its 
value is $\simeq 2$
\footnote{This is true until the constraint (\ref{upper}) on the mixing angle is
adopted, otherwise $y_c$ can grow to much higher values \cite{fv1,dll} as an effect of 
sterile neutrino production. The asymmetry generation should go off when $y_c$ becomes
so high to be in the tail of the distribution ($y_c> y_c^{\rm off}\gtrsim 10$), 
but this has not yet been studied by numerical calculations and 
the exact value of $y_c^{\rm off}$ has not been determined.}.
The {\em total effective asymmetry} $L^{(\alpha)}$ is defined as
\footnote{When not necessary we will drop the subscript in $L^{(\alpha)},\tilde{L}^{(\alpha)}$
and just indicate them with $L,\tilde{L}$.}:
\begin{equation}
L^{(\alpha)} \equiv 2\,L_{\nu_{\alpha}}+\tilde{L}^{(\alpha)}
\end{equation}
where $\tilde{L}$ contains the asymmetry of 
the other neutrino flavours ($\beta \neq \alpha$)
and also a contribution coming from electrons and baryons. 
We assume that $\tilde{L}$ is of the same
order of magnitude as the baryon asymmetry $ \sim 5 \times 10^{-10}$,
a value that we adopted in our calculations
\footnote{In any case the final value of $L$ is not dependent on 
a particular choice for $\tilde{L}$. Moreover note that in the 
realistic case of multiflavor oscillations a mixing 
$\nu_{\beta\neq\alpha}\leftrightarrow \nu_s$
could create an $|L_{\nu_{\beta}}|\gg 5\times 10^{-10}$ prior the
occurrence of $\nu_{\alpha}\leftrightarrow \nu_s$ oscillations. 
Thus, in the case of multiflavor mixing, our assumption
is valid if either the $\nu_{\alpha} \leftrightarrow \nu_s$ oscillations 
have the largest $|\delta m^2|$ or the oscillations $\nu_{\beta}\leftrightarrow \nu_s$
do not have the right parameters (negative $\delta m^2$, large enough mixing angle, \dots )
to create a large $|L_{{\nu}_{\beta}}|$.}.
The evolution of the total asymmetry experiences different stages.
At high $T \stackrel{>}{\sim} T_c$, oscillations 
drive $|L| << |\tilde{L}|$. When $T$ approaches $T_c$ 
a decrease of $|L|$ to arbitrarly small values is prevented
by the presence of chemical potentials in neutrino equilibrium distributions
and $|L|$ starts to grow slowly.
At $T=T_c$, by definition, $L=\tilde{L}$ and $L_{\nu_{\alpha}}=0$
\footnote{Within the static approximation a definition of $T_c$ is more rigorous
\cite{ftv,fv1}. This is a sort of numerical definition.}. 
Until this momentum, the resonant momentum $y_{\rm res}=p_{\rm res}/T$ of neutrinos and antineutrinos is practically the same and evolving as $\propto T^{-3}$.
At $T\lesssim T_c$ the asymmetry growth becomes exponential and 
the neutrino and antineutrino resonances quickly become separated: for $L>0$ 
the antineutrino resonant momentum gets smaller 
than the critical momentum $y_c$,
while the neutrino resonant momentum increases rapidly moving into
the tail of the distribution where the number of
neutrinos becomes negligibly small (and vice-versa for negative $L$).
When $|L^{(\alpha)}|$ reaches a threshold value 
$L^{(\alpha)}_t \simeq 1.0(0.4)\times 10^{-6}\,(y_c\,|\delta m^2|/{\rm eV}^2)^{1/3}$
for $\alpha=e,(\mu,\tau)$, non linear effects of the medium start to
act and the exponential growth turns into a power law (with
$L \propto T^{-4}$ initially), while (for $L$ positive) anti-neutrino resonant 
momentum gets a minimum value $y_{\rm res}^{\rm min}$ and then starts 
to grow again.
A relation between $y_{\rm res}$ and $L$ can be obtained from
the resonance condition and is given by
the following approximate expression: 
\begin{equation}\label{yres}
y_{\rm res}\simeq {|\delta m^2|/{\rm eV}^2 \over a_0\,L\,(T/MeV)^4},
\end{equation} 
with $a_0\equiv (4\,\sqrt{2}\,\zeta(3)/\pi^2)\times 10^{12}\,G_F\,{\rm MeV}^2\simeq 8$.
When $y_{\rm res}\sim 10$ the asymmetry stops its growth and gets 
frozen to its final value $L_{\rm f}\sim 0.3$. It is simple, 
from the Eq. (\ref{yres}), to see that this happens at
a temperature given by:
\begin{equation}\label{Tf}
T_f \approx 0.5\,{\rm MeV}\,\left( |\delta m^2|\over {\rm eV}^2 \right)^{1\over 4}.
\end{equation}

It is quite interesting that this complicated behaviour of the neutrino
evolution can be regarded as the effect of two different physical regimes.
The relevant quantity is the ratio 
$r\equiv ({\ell}_{\rm int}/{\ell}_{\rm osc})_{\rm res}$ betweeen the interaction
length and the oscillation length for the resonant neutrinos. 
If $r\lesssim 1$ collisions are able to inhibit the MSW effect and the regime
is {\em collision dominated} and can be described by a simplified set of 
kinetic equations obtainable within the static approximation procedure \cite{fv1,bvw}.
On the other hand if $r\gtrsim 1$, the MSW effect can take place and the regime
is MSW {\em dominated}. In \cite{comment} it has been clarified that the 
transition between the two regimes depends on the value of the asymmetry 
and not just on the temperature. It is in fact possible to show that the 
following asymptotic expressions for $r$ holds:
\begin{eqnarray} 
r & \simeq & {\sin 2\theta_0\over 0.8(2.0)10^{-2}}\hspace{5mm} (|L|\ll L_t) \\ 
r & \simeq & {\sin 2\theta_0\over 0.8(2.0)10^{-2}}\,\left({|L|\over L_t}\right)^{3\over 2}
  \hspace{5mm} (|L|\gg L_t)  \label{r}
\end{eqnarray}
First of all note that a collisional regime can exist only if  
$\sin 2\theta_0\lesssim 10^{-2}$. Thus $\sin 2\theta_0 \simeq 10^{-2}$ 
is a border between two very different regions of mixing parameters. 
In this paper we will only be interested
to the case of {\em very small mixing angles} ($\sin 2\theta_0\ll 10^{-2}$) 
in which collisions are important
at low values of the asymmetry ($|L|<L_t$), independently on the temperature. In this case
only when $|L|\gtrsim (10^{-2}/\sin 2\theta_0)^{2/3}\,L_t$ the MSW regime can begin. 
It is then crucial that, during the collisional regime, a fast growth of the asymmetry 
occurs for the MSW effect to start to be effective 
and for the resonance of neutrinos and antineutrinos to become separated. 
If in fact now one assumes (for $L>0$) that the expansion is slow enough that the 
resonant momentum of antineutrinos, after the minimum, increases so slowly that antineutrinos
adiabatically cross the resonance, then all $\alpha$-antineutrinos 
are converted into 
antisterile neutrinos. This easily explains why the final values of the $\alpha$-neutrino asymmetry  are in a range of values slightly lower than the value
$N_{\bar{\nu}_{\alpha}}(T_{\rm in})/N_{\gamma}(T_{\rm in})=0.375$, 
corresponding to an extreme situation in which all anti-neutrinos 
are converted. This simple picture is the {\em adiabatic approximation} 
first described in \cite{fv2}. 

As we said,  the numerical solution of the QKE's have shown that for 
$|\delta m^2|\geq 10^{-4}\,{\rm eV}^2$ and for the mixing angle 
within the limits (\ref{lower}) and (\ref{upper}) (without however a systematic
investigation at the borders) this physical approximated two-regime picture works very well 
in the description of the neutrino asymmetry evolution and of its final value
\footnote{Excluding the phenomenon of rapid oscillations at the onset of the
exponential growth for $\sin^2 2\theta_0\gtrsim 10^{-6}$ \cite{ropa} that requires
some further physical picture still not completely understood.}.

{\em What about for $|\delta m^2|< 10^{-4}\,{\rm eV}^2$ ?} 
{\em Does neutrino asymmetry generation still occur ?}
{\em Does the static-adiabatic approximation picture still work ?}

 The authors of ref. \cite{bar}, with an analytical procedure based
on the mean momentum approximation
\footnote{This approximation works very well in the description
of the sterile neutrino production \cite{kimmo} for reasons 
explained in \cite{dll}. However in the case of the neutrino asymmetry
generation, especially in the MSW dominated regime, the momentum dependence
is such a crucial feature that the mean momentum approximation 
is a very drastic one and in our opinion it should 
be regarded more as a toy model.}, were
finding that, even though an exponential growth occurs 
for $|\delta m^2|\gtrsim 10^{-7}\,{\rm eV}^2$, soon 
the non linear term in the effective potential 
is able stop the growth and concluded that
no neutrino asymmetry (larger than $10^{-7}$) can be generated in any point 
of the parameter space.  This result seemed to be confirmed
by numerical results \cite{ekm}. These 
were obtained also employing the mean momentum approximation
for $\delta m^2=-10^{-5} {\rm eV}^2$ and $\theta_0=0.1$.
At some critical temperature an  oscillatory behaviour 
of the neutrino asymmetry, with an amplitude exponentially growing,
was found, but the amplitude was not growing higher than $10^{-7}$,
although this may simply be because the evolution was stopped at 
$T\simeq 0.3 {\rm MeV}$, below the freezing temperature of 
neutron to proton abundance ratio or simply an artifact of not
including the momentum degree of freedom. 
In any case the authors of Ref.\cite{ekm} concluded that
neutrino oscillations cannot generate a large neutrino asymmetry in
the early Universe, in apparent agreement with \cite{bar}. 
In \cite{ftv,fv1,shi,fv2} it was shown that a large neutrino asymmetry 
can be generated due to the
crucial role of the collisions, thus disproving the general conclusions 
of \cite{bar,ekm}, although this result is still not necessarily in disagreement 
with the conclusions of \cite{bar,ekm} for $|\delta m^2|\lesssim 10^{-4}$ 
and this point has still remained unclear.

It has also to be said that the
main attention was put at higher values of $|\delta m^2|$ because 
only in this case the asymmetry generation occurs early enough to modify
the standard BBN predictions for the nuclear abundances. Moreover there are also 
qualitative reasons to suspect that at low $|\delta m^2|$ 
the static approximation is not valid.
 In \cite{ftv,fv1,bfv} it has been infact pointed out that for 
$|\delta m^2|\lesssim 10^{-4}{\rm} eV^2$, oscillations betweeen collisions can 
give a non negligible contribution to the rate of the neutrino asymmetry, 
suggesting a possible deviation from 
the static approximation. This statement
is justified observing that, if one requires 
that the change of the mixing angle in matter is negligible between collisions
on average
\footnote{This is equivalent to imposing that the derivative of the 
effective potential, with negligible neutrino asymmetry, is much less
that the total interaction rate}, 
then the critical temperature has to be
higher than approximately $3\, {\rm MeV}$ and from the Eq. (\ref{Tc}) 
this implies just $|\delta m^2|\gtrsim 10^{-4} {\rm eV}^2$.
On the other hand we already noted that the role of collisions in inhibiting 
the MSW effect, for low values of the asymmetry, 
is effective at all temperatures (for $\sin 2\theta_0 \lesssim 10^{-2})$.

Which kind of deviations from the static
approximation are then expected at small $|\delta m^2|$ and, above all, does the asymmetry generation still occur ?  Are oscillations in the neutrino asymmetry expected to appear as the mean momentum calculations of \cite{ekm} suggest ?

The purpose of this paper is to answer these questions, mainly 
understanding the exact border of neutrino asymmetry generation at low values
of $|\delta m^2|$. We will however also give some results concerning the presence of 
neutrino asymmetry oscillations. Moreover we will study the border of the generation  
at small mixing angles using the QKE's and show that the limit (\ref{lower}), 
derived in the static approximation, has actually to be replaced with a less
stringent one (section II). We will show that the small mixing angle border can
be explained within an analysis on the limits of validity of the 
adiabaticity approximation in the description of the MSW dominated regime. 
This analysis will also involve
extending the adiabatic approximation in the border region and outside 
by means
of the  Landau-Zener approximation, that is able to describe how the
neutrino asymmetry generation dies at small mixing angles (section III).  
In section IV we will study the cosmological consequences of the neutrino asymmetry 
generation in the light of the new results. If neutrino asymmetry generation
occurs also at low values of $|\delta m^2|$, even though this  implies that it is generated
at temperatures too low to produce remarkable changes in standard BBN,
one can still consider which kind  of cosmological effects (and maybe observations !) 
can result from it. In particular
we will correct some wrong claims, which have 
appeared recently in the literature, 
on the role that the neutrino asymmetry generated in active-sterile neutrino 
oscillations would have on CMBR. We will consider the implications on the
possibilities of a detection of relic neutrino background by the light of recent studies \cite{smirnov}. In section V we will draw the 
conclusions of this work.

\section{Numerical results}
\subsection{Set of equations and numerical procedure}

We have numerically solved the QKE's
for the density matrix distribution in the momentum space, 
describing the statistical properties of the $\nu_\alpha, \nu_s$ states,
including their mixing. Since we are interested in extending previous
studies to values of $|\delta m^2|/{\rm eV}^2 < 10^{-4}$ and since from the
equation (\ref{Tf}), obtained within the adiabatic approximation,
one would have that the neutrino asymmetry is generated down to 
temperatures below $T\sim 0.05\,{\rm MeV}$, we have to take into account
the occurance of electron-positron annihilations that will make $T_{\nu}\neq T$.
The first modification is that the re-scaled momentum 
$p\,R/R_{\rm in}=p\,T_{\rm in}/T_{\nu}$ 
and thus now all neutrino distributions are more conveniently expressed 
in terms of $y\equiv p/T_{\nu}$.

The QKE's describing neutrino propagation are given by
the following set of equations for 8 distributions in momentum 
space\cite{QKE},
$\vec{P}(y,t), P_0(y,t), \vec{\bar{P}}(y,t), \bar{P}_0(y,t)$
(for notational simplicity we drop the dependence on momentum and on time):
\begin{eqnarray}
{dP_x\over dt} & = & -\lambda\,P_y-D\,P_x            \label{dPx} \\
{dP_y\over dt} & = & \lambda\,P_x-\beta\,P_z-D\,P_y  \label{dPy} \\
{dP_z\over dt} & = & \beta\,P_y+R                   \label{dPz} \\ 
{dP_0\over dt} & = & R                              \label{dP0}
\end{eqnarray}
The function $R(y,t)$ takes into account the {\em (thermal) redistribution} of
neutrinos at each quantum state due to the effect of collisions, that in a linear
approximation is given by \cite{f98,bvw}:
\begin{equation}\label{R}
R \simeq \Gamma\,
\left[{f_{\xi}^{\rm eq}\over f_0}-{1\over 2}\,(P_0+P_z)\right]
\end{equation}
where $f^{\rm eq}_{\xi}(y,t)\equiv [1+e^{y-\xi(t)}]^{-1}$ is the Fermi-Dirac distribution, 
$\xi(t)$ is the {\em (dimensionless) chemical potential} and $f_0\equiv f^{\rm eq}_{\xi =0}$. 
The variables $P_0$  and $P_z$ can be replaced with
the variables $z_{s}= (P_0 - P_z)/2$ and $z_{\alpha}=(P_0+P_z)/2$. 
These variables are the diagonal entries of the density matrix and 
are related to the $\alpha$ and sterile neutrino distributions simply by
$z_{\alpha,s}\equiv f_{\nu_{\alpha},s}/f_{0}$. In this case the
equations (\ref{dPz}) and (\ref{dP0}) are replaced by the following ones:
\begin{eqnarray}
{dz_s\over dt} & = & -{1\over 2}\beta\,P_y \\
{dz_{\alpha}\over dt} & = & -{dz_s\over dt}+R
\end{eqnarray}
while the repopulation function $R$ now can be recasted as 
$\,R=\Gamma\,(z_{\alpha}^{\rm eq}-z_{\alpha})$.

The anti-neutrinos barred distributions $\vec{\bar{P}},\bar{P}_0$ obey 
the same equations with the same {\em total collision rate} 
$\Gamma$, the same {\em decoherence parameter} $D$,  
and the same function $\beta$
\footnote{We neglect a small term, proportional to the asymmetry $L$, 
that makes the collision rate $\Gamma$ for neutrinos 
and $\bar{\Gamma}$ for anti-neutrinos different \cite{bfv}. Within
the static approximation this term proves not to give any difference. It has 
to be investigated whether in the region of rapid oscillations this term can play or 
not any role.}. 
The function $\lambda$, containing the effective potential of neutrinos, depends
explicitly also on $L$ and for anti-neutrinos it changes such that 
$\bar{\gamma}(L)=\gamma(-L)$.
Thus the presence of a {\em total asymmetry} $L$ couples together 
the neutrino and antineutrino sets of equations. 
They are coupled also by the presence of the chemical potentials in 
the equilibrium active neutrino distributions. These are infact different 
if a {\em neutrino asymmetry} is present, considering that:
\begin{equation}\label{asy} 
L_{\nu_{\alpha}} = {1 \over 4\zeta (3)}\int^{\infty}_0\,dy\,y^2\, 
\left[{1\over 1 + e^{y-\xi}} - {1 \over 1 + e^{y-\bar{\xi}}}\right]  
\end{equation}

Note that if $\tilde{L}=0$, then one would have simply 
$L=2\,L_{\nu_{\alpha}}$ and $L=L_{\nu_{\alpha}}=0$ would be an exact solution of the
set of equations. Above the critical temperature this would be a stable solution and
$L$ would get so small that the statistical fluctuations on very small scales 
would determine the sign of the asymmetry below the critical temperature when the solution becomes unstable. This sign would be different on distances as small as the
scale of the dominant statistical fluctuations that is smaller than the interaction length \cite{ropa}. On these scales neutrino free-steam and the free-streaming is faster than
the neutrino asymmetry generation 
such that immmediately regions with different sign
would destroy each other and no asymmetry generation could take place. This picture is  
consistent with the idea that the generation of a neutrino asymmetry can occur 
only if CP is violated at some level,
in this case due to the 
presence of a CP asymmetric medium ($\tilde{L}\neq 0$) that can
push the neutrino asymmetry toward some direction in any point of the space 
and strongly enough to dominate on statistical fluctuations. This surely happens in a large region of mixing parameters where the presence of 
rapid oscillations can be excluded \cite{ropa}.

Note also that the integral equation (\ref{asy}) is coupled 
to the set of differential equations for the $\vec{P},P_0,\vec{\bar{P}},\bar{P}_0$, 
as the value of the neutrino asymmetry appears in the expression for $\lambda$.
However, as already discussed in previous works \cite{fv1,fv2,foot,ropa},
it is numerically more convenient to calculate the neutrino asymmetry
not from the equation (\ref{asy}) but through the 
following integro-differential equation:
\begin{equation}\label{dL}
{dL_{\nu_\alpha} \over dt} =
{1 \over 8\zeta (3)}\int
\beta(y) [P_y (y) - \bar P_y(y)] f_0(y)\,y^{2}dy.
\end{equation}
The set of equations (\ref{dPx}-\ref{R},\ref{dL}) is still not `closed' as one still
needs some equations describing the evolution of the chemical potentials
$\xi$ and $\bar{\xi}$ in the equilibrium active neutrino 
distributions. We use the instantaneous chemical decoupling
procedure developed in Ref.\cite{bfv}
with kinetic decoupling temperatures,
$T_{\rm dec}^{\alpha} = 3.5 $ MeV  for 
$\alpha = \mu,\tau$ and $2.5$ MeV for $\alpha = e$. 
The approximate validity of this procedure has been
checked in Ref.\cite{foot}. 

The last step is to replace the time with temperature. 
It is well known that the neutrino temperature $T_{\nu}$ can be expressed as a function  of the
photon temperature $T$, simply imposing entropy conservation 
during electron-positron annihilations:
\begin{equation}
T_{\nu}=T\,\left[{g_S(m_e/T)\over g_S(0)}\right]^{1\over 3}
\end{equation}
with the effective number of (entropy) degrees of freedom of photons and electron-positrons
given by:
\begin{equation}
g_S(x)=2+{45\over\pi^4}\,\int_0^{\infty} dy\,y^2
\frac{\sqrt{y^{2}+x^{2}}+\frac{1}{3}\frac{y^2}{\sqrt{y^{2}+x^{2}}}}
{e^{\sqrt{y^{2}+x^{2}}}+1}
\end{equation}
and $g_S(0)=11/2$ (in {\bf figure 1} both $g_s$ and $T_{\nu}/T$ are shown for
illustrative purposes). One can choose either $T$ or $T_{\nu}$ as the 
independent variable for the numerical calculations (we used both choices 
in two different codes). For our discussion it is simpler to choose $T_{\nu}$ 
and thus we have to express all quantities as a function of $T_{\nu}$. 
The quantity $\beta$ can be written as:
\begin{equation}
\beta(y,T_{\nu})={\delta m^2\over 2\,y\,T_{\nu}}\,\sin 2\theta_0
\end{equation}
Since we treat neutrinos as fully decoupled during the period of electron-positron
annihilations, interactions will be effective only
until $T\simeq T_{\nu}$ and thus we can simply write:
\begin{equation}
\Gamma(y,T_{\nu})\simeq 1.27\,(0.92)\,G^2_F\,T_{\nu}^5\,y 
\;\;\;\;\;\;\;\;\;\alpha=e\,(\mu,\tau)
\end{equation}
The quantity $\lambda$ can be written as:
\begin{equation}
\lambda(y,T_{\nu},L)=-{\delta m^2 \over 2\,y\,T_{\nu}}
                    [\cos2\theta_0 - b(y,T_{\nu})\pm a(y,T_{\nu},L)].
\end{equation}
where the $+$ ($-$) sign holds for neutrinos (anti-neutrinos).
The term $b(y,T_{\nu})$ is the finite temperature contribution
to the (dimensionless) effective potential in the early Universe \cite{Notzold}.
This term is important when $|L|\ll L_t$, but it becomes negligible after
the exponential growth, when $|L|\gg L_t$, for $T\lesssim T_c$. 
We will mainly consider $T_c \gtrsim 0.25 {\rm}\,{\rm MeV}$
for which electron-positron number densities are still 
approximately equal to the ultra-relativistic limit.
Thus we can just replace $T$ with $T_{\nu}$ and write
\footnote{Actually we will even consider values of 
$-\delta m^2/{\rm eV}^2\lesssim 10^{-11}$ for which $T_c\lesssim 0.25\,{\rm MeV}$.
In this case one should replace the constant $T_{\alpha}$ with a function
of temperature taking into account that electron and positrons
number densities decrease during annihilations. We
neglected this modification and used the standard
Notzold-Raffelt expression for the finite temperature term.}:
\begin{equation}\label{bterm}
b(y,T_{\nu})= 
-{\rm eV^2\over \delta m^2}\left(T_{\nu}\over T_{\alpha}\right)^6 \,y^2
\end{equation}
where $T_{\alpha}\simeq 18.9\,(23.4){\rm MeV}$ for $\alpha=e\,(\mu,\tau)$.
The term $a$ is proportional to the total asymmetry $L$. 
Electron-positron annihilations do not change electron asymmetry in $\tilde{L}$ 
and as usually is done, we also neglect a small change in $\tilde{L}$ 
arising from neutron-proton conversions. Thus we can write the term $a$
in the usual way, where now however $T$ has to be replaced with $T_{\nu}$
because of the definition (\ref{LX}) of particle asymmetry that we are using:
\begin{equation}\label{aterm}
a\simeq -\,a_0\,{\rm{eV^2}\over \delta m^2}\,L\,T_{\nu}^4\,y
\end{equation} 
So far, the
electron positron-annihilations have not caused any change in the
equations when these are written in $T_{\nu}$. This because we could approximately assume that $\alpha$- neutrinos, during the electron positron annihilations, are not 
sensitive to all other particle species, either because 
they do not interact with them
any more or because the contribution of the other particle species in the effective potentials
can be neglected when $T\lesssim 0.25\,{\rm MeV}$ and $|L|\gg L_t$. There is however 
an indirect effect due to the expansion rate. In replacing the time $t$ with the 
neutrino temperature $T_{\nu}$, one needs to calculate the quantity:
\begin{equation}\label{dTdt}
{dT_{\nu}\over dt} \simeq -{1.66\,\sqrt{g_{\rho}}\,T^{3}_{\nu}\over M_{\rm Pl}}
\,\left({T\over T_{\nu}}\right)^2,
\end{equation}
where $M_{\rm Pl}$ is the Planck mass and
the number of (energy density) degrees of freedom is given by
\footnote{We neglect any contribution to $g_{\rho}$ from the sterile neutrino
production or from the modification of $\alpha$ neutrino distribution, due
to the generation of the asymmetry. This because we are primarily
interested in small 
values of mixing angles satisfying the Eq. (\ref{upper}) and to values
of $|\delta m^2|/{\rm eV}^2 \ll 100$ for which most of the neutrino asymmetry is
generated below $T_{\rm cdec}^{\alpha}$ and thus $\Delta N^{\rho}_{\nu}\simeq 0$.}:
\begin{equation}
g_{\rho}(T_{\nu})=2+{21\over 4}\,\left({T_{\nu}\over T}\right)^4+
{60\over \pi^4}\,\int_0^{\infty}\,dy\,{y^2\,\sqrt{y^2+x^2}\over 1+e^{\sqrt{y^{2}+x^{2}}}}
\end{equation} 
with $x\equiv m_e/T$ and where $T$ has to be regarded as a function of $T_{\nu}$
(in figure 1 the function $g_{\rho}$ is also plotted).
Thus electron-positron annihilations cause a change in the 
expansion rate. In section IV we will obtain
physical insight into this effect.

In {\bf figures 2,3,4} we show the evolution of the absolute value of 
the total asymmetry  with the temperature $T$. In {\bf figure 2} 
we show solutions for many different values of $-\delta m^2$ but for a fixed
value of the mixing angle $\sin ^2 2\theta_0=10^{-8}$ and
for $\alpha=e$. The solid lines are the solutions
obtained solving the set of QKE equations.
The dotted lines are curves
obtained integrating the QKE's only until $r<3$ (see the Eq. \ref{r}). 
From that moment we started to use the {\em adiabatic approximation} as described in \cite{fv2,foot}. It can be seen how the asymmetry generation
occurs at values of $-\delta m^2$ much below $10^{-4}\,{\rm}\,{\rm eV}^2$ 
and goes off only around $-\delta m^2\simeq 10^{-10}{\rm eV}^2$. In particular
for $-\delta m^2=10^{-10}\,{\rm eV}^2$ the final value of neutrino
asymmetry  is still at the level of $10^{-5}$.

The numerical convergence of the solutions has been obtained integrating 
around the resonance \cite{foot,ropa} but the interval of the integration 
highly increases for low $|\delta m^2|$ and covers up to two orders of magnitude 
in momentum space. The behaviour at low values of the asymmetry is well 
reproduced by the static approximation at any value of $|\delta m^2|$. 
In particular note that no sign changes are present. 

In {\bf figure 3} solutions for $\sin^2 2\theta_0=10^{-7}$,  
$-\delta m^2/{\rm eV}^2=100,1,10^{-2},10^{-8}$ and $\alpha=e$ are shown. 
While for the first two cases there are no sign changes, which was already
known\cite{ropa},
in the case $-\delta m^2=10^{-8}$ two sign changes appear. Note that, changing the
initial neutrino asymmetry, the solutions converge to the same values at some stage.
The final sign is still the same of $\tilde{L}$, like the static approximation predicts. For the numerical convergence of this solution a momentum integration over an interval of 5 orders of magnitude has been necessary, much larger than for $\sin^2 2\theta_0=10^{-8}$. 
In {\bf figure 4} another example
is shown with the mixing parameters $\sin^2 2\theta_0 = 10^{-7}$,
$\delta m^2/eV^2 = -10^{-7}$ for the $\alpha=\mu$ case.
In this case three sign changes are present and the final sign is therefore
opposite to that of $\tilde{L}$. We tested this solution with 
both codes, obtaining the same behaviour.
Thus we can say that the numerical calculations suggest the appearance
of {\em slow oscillations} in the neutrino asymmetry. These seem to have a different
nature than the rapid oscillations observed in \cite{ropa} for $\sin^2 2\theta_0\gtrsim 10^{-6}$
and $10^{-2}\lesssim |\delta m^2|/{\rm eV}^{2}\lesssim 500$. They are `slower' and with
just a few sign changes and
do not occur during the exponential growth but before, above
the critical temperature. Moreover, even though the integration has to be very accurate,
a numerical convergence can be obtained and thus any character of chaoticity seems to be excluded. This is also supported by the fact that 
starting with two slightly different
initial conditions the final sign is the same. We can thus exclude the possibility
that statistical fluctuations, estimated in \cite{ropa}, can influence the solutions. 
However, it would be desirable to have some analytical insight supporting the
presence of sign changes found in the numerical results. In any case, if confirmed,
they would suggest a deviation from the static approximation at 
low $|\delta m^2|$ and sufficiently high mixing angles ($\sin^2 2\theta_0 \sim 10^{-7}$)
due to the presence of sign changes that are not predicted by the static approximation.
We did not investigate whether at even higher mixing angles, rapid oscillations appear,
as in the case of high $-\delta m^2$. This requires a dedicated analysis
beyond the goals of this work.

In {\bf figures 5a} we give the final value of the asymmetry for $\alpha=e$, 
five values of the mixing angle 
($\sin^2 2\theta_0= 10^{-7}$, $10^{-8}$, $10^{-9}$, $4\times 10^{-10}$, $10^{-10}$) 
and for $-\delta m^2$ in the range $10^{-13} < -\delta m^2/eV^2 < 10^4$
\footnote{The chosen upper limit for $|\delta m^2|$ comes from the requirement to have
the critical temperature $T_c\lesssim 100\,{\rm MeV}$ considering that, for
higher temperatures, 
the expression that we used for the effective potentials are not valid.
However in the case $\alpha=e$, direct mass measurements
from Tritium $\beta$ decay limit $-\delta m^2$ to be less than 
$\simeq 100\,{\rm eV^2}$.}. 
In {\bf figure 5b} the final value of the asymmetry is
shown in the case $\alpha=\mu$ and for two values of the mixing angle
($\sin^2 2\theta_0= 10^{-7}$ and $10^{-8}$) and one can see that there is not
much difference with the case $\alpha=e$. This because in the MSW regime, when
the neutrino asymmetry gets its final value, the finite temperature term in
the effective potential (the $b$-term, see Eq. (\ref{bterm}) ) 
is negligible and only the Wolfenstein term is 
relevant (the $a$-term, see Eq. (\ref{aterm}) ).
This term is the same for $\alpha=e$ and $\alpha=\mu$.

For values of the mixing angles $\sin^2 2\theta_0= 10^{-7}$, $10^{-8}$ and $10^{-9}$, 
the final value of the neutrino asymmetry can be quite easily understood as being due
to MSW transitions as the resonance momentum passes through the neutrino distribution.
Previous work focussed on the adiabatic limit where complete
conversion occured. The solid curves describe the results obtained solving
the QKE. The dotted lines correspond to the results obtained with the adiabatic approximation,
with a slight difference between figure 5a and figure 5b. In figure 5a ($\alpha=e$), 
the results correspond to the solutions in figure 2 and 3 obtained combining QKE and
the adiabatic approximation as already illustrated. In figure 5b the adiabatic approximation
is started from $T\sim T_c/2$ with an initial asymmetry corresponding to a resonant
momentum $y_{\rm res}^{\rm min}=0.3$ (see Eq. \ref{yres}). 
In the first case one gets a compromise
of a fast procedure (most of the CPU time is required in the MSW dominated regime) but
still able to reproduce accurately the 
correct neutrino asymmetry evolution at any temperature.
In the second case one has a super-fast procedure able to reproduce 
accurately the final stages ($|L| \gtrsim 10^{-2}$) of the evolution,
and a fair description of the intermediate stages.
At high values of $-\delta m^2$ the adiabatic approximation well describes the
results obtained using the QKE's. Note also that, for 
$|\delta m^2|/{\rm eV}^2\gg 10^{-2}$, not all anti-neutrinos are converted into
anti-sterile neutrinos (we refer to positive $L$, otherwise neutrinos would be converted
instead of anti-neutrinos) and $|L_{\nu_{\alpha}}^{\rm fin}|$ can be well below the 
value $0.375$ corresponding to a full conversion and for $y_{\rm res}^{\rm min}\ll 1$. 
This is because at high values of $|\delta m^2|$ the MSW dominated regime occurs mostly 
at temperature above $\sim 1\,{\rm MeV}$ and 
collisions are able to re-distribute thermally the 
asymmetry after it has
been produced at $y=y_{\rm res}$.
In particular this means that (on statistical average) some anti-neutrinos in quantum states 
with $y>y_{\rm res}$ will move to quantum states with
$y<y_{\rm res}$. Since $y_{\rm res}$ increases with time, 
this means that these anti-neutrinos
cannot be converted any more. In a pictorial way, we can say that 
these anti-neutrinos {\em evade} the resonance and manage not to be converted. 
This {\em evasion effect} can be described correctly only numerically.

For lower values of $-\delta m^2$ the asymmetry becomes
large only for 
low temperatures $\stackrel{<}{\sim} 1\,{\rm MeV}$, where collisions
have an approximately negligible effect and the evasion effect does not occur.
The final value of the neutrino asymmetry gets very close to 
the upper limit of $0.375$ around 
$-\delta m^2\sim 10^{-2}\,{\rm eV}^2$. For lower values of $-\delta m^2$ the adiabatic approximation predicts that the final value remains approximately 
constant at $0.375$ 
\footnote{More precisely, as it is shown figure 5a,
for $\sin^2 2\theta_0=10^{-8}$ in the adiabatic approximation, the final value gets again
slowly  lower than $0.375$ for $-\delta m^2/{\rm eV}^2\simeq 10^{-2}$. 
This because in that case we integrated the QKE's in the collision dominated regime and 
in this case the correct value for the initial $y_{\rm res}^{\rm min}$ at which the MSW conversions start is reproduced. This value gets as high as 1 at $-\delta m^2=10^{-8}\,{\rm eV}^2$ and thus not all-antineutrinos are converted as it happens in figure 5b. 
On the other hand, for 
larger $\sin^2 2\theta_0$, the value of $y_{\rm res}^{\rm min}\ll 1$
and the two procedures, the fast one and the super-fast one, give the same answer.}.
However at low enough values of  $-\delta m^2$, the adiabatic approximation clearly fails in reproducing the results obtained fully integrating the QKE's. These show that, for a fixed value of $\sin^2\,2\theta_0$, there is a value of $-\delta m^2/{\rm eV}^2$ below which the final value of the asymmetry becomes rapidly negligible. The region of mixing parameters where
the adiabatic approximation holds and the final asymmetry 
$|L_{\nu_{\alpha}}^{\rm fin}|\gtrsim 0.8\,|L_{\nu_{\alpha}}^{\rm fin}|^{(\rm ad)}$, 
with the value of the adiabatic limit 
$|L_{\nu_{\alpha}}^{\rm fin}|^{(\rm ad)}$ in the range $0.26-0.375$ (the exact value 
depending on $-\delta m^2$), is given approximately by:
\begin{equation}\label{ad}
\sin^2 2\theta_0\,\left(|\delta m^2|\over{\rm eV}^2\right)^{{1\over 4}}
 \stackrel{>}{\sim} 1.2\times 10^{-9}
\end{equation}
It is also simple to infer from figure 5a
\footnote{We concentrated, for values of the final asymmetry much less than $0.01$,
on the case $\alpha=e$. However in the case $\alpha=\mu$ one does not expect
differences, as the Wolfenstein term in the effective potential,
responsible for the final value of the asymmetry in the MSW regime,
is the same.} the following
equations describing the iso-$|L_{\nu_{\alpha}}^{\rm fin}|$ curves in the space 
of mixing parameters:
\begin{equation}\label{iso}
\sin^2 2\theta_0\,\left(|\delta m^2|\over{\rm eV}^2\right)^{{1\over 4}}
\simeq 0.8\times 10^{-9}\, |L_{\nu_{\alpha}}^{\rm fin}|^{1\over 4}
\end{equation}
For $|L_{\nu_{\alpha}}^{\rm fin}|>10^{-5}$ one finds a larger region compared to
the  condition (\ref{lower}) from the static approximation. This is 
because the MSW effect enhances the neutrino asymmetry generation compared 
to a collision dominated regime. The curves (\ref{iso}) are however valid, 
if one consider values of $|L_{\nu_{\alpha}}^{\rm fin}|\geq 10^{-5}$, only for
$\sin^2 2\theta_0\gtrsim 10^{-9}$. 
For smaller angles the validity of (\ref{iso}) breaks down for 
$10^{-5}\leq |L_{\nu_{\alpha}}^{\rm fin}|\lesssim 
L_{\nu_{\alpha}}^{\star}(\sin^2 2\theta_0)$.
For example for $\sin^2 2\theta_0=4\times 10^{-10}$ we found that 
$L_{\nu_{\alpha}}^{\star}\simeq 10^{-3}$ corresponding to 
$-\delta m^2/{\rm eV^2}=10^{-2}$ and this is clearly visible
in {\bf figure 5a}.  For lower values of $-\delta m^2$ 
and a fixed value of $\sin^2 2\theta_0$, there is a sharp
cutoff in the generation. 

In {\bf figure 6} we plotted the iso-$|L_{\nu_{e}}^{\rm fin}|$ curves
for $-\log |L_{\nu_{e}}^{\rm fin}|=1,2,3,4,5$. 
For $\sin^2\,2\theta_0\gtrsim 10^{-9}$ the curves (\ref{iso}) are found 
while for smaller mixing angles they do not hold and the generation 
goes off much faster for a decreasing $-\delta m^2$ and for a fixed 
value of $\sin^2 2\theta_0$.

One can easily explain this cutoff in the generation of neutrino asymmetry at low
values of mixing angles ($\sin^2 2\theta_0\lesssim 10^{-9}$) where 
Eq.'s (\ref{iso}) start not to be valid for $|L_{\nu_{e}}^{\rm fin}|<L^{\star}$.
This can be done in terms of current understanding of neutrino asymmetry generation
outlined in the introduction. In fact for too small mixing angles the collision  
dominated regime
\footnote{Note that the rate of neutrino asymmetry generation
in the collision dominated regime, as deduced within the static approximation,
is proportional to $\sin^2 2\theta_0$ \cite{ftv,fv1}.} 
is not efficient enough in generating a neutrino asymmetry 
$|L^{(e)}|\gtrsim (10^{-2}/\sin 2\theta_0)^{2/3}\,L_t$ 
and the MSW dominated regime cannot start at all (see Eq. (\ref{r}) ).  
This can be seen in {\bf figure 7}, where we plotted
the evolution of the neutrino asymmetry with temperature for
$\sin^{2}2\theta_0=4\times 10^{-10}$ and a few values of $-\delta m^2/{\rm eV}^2$.
One can see how, in the case $-\delta m^2/{\rm eV}^2= 10^{-3}$, the asymmetry
gets frozen much before it can get the value 
$(10^{-2}/\sin 2\theta_0)^{2/3}\,L_t\simeq 10^{-4}$ and the MSW 
dominated regime never starts.
 
We are still lacking in a physical description of the (\ref{iso}) themselves
for $\sin^2 2\theta_0\gtrsim 10^{-9}$. In the next section, we will 
develop a simple physical picture providing such a description improving
the adiabatic approximation in the MSW dominated regime.

\section{The Landau-Zener approximation}

We will now develop a useful extension to the adiabatic approximation
by taking into account non-adiabatic effects which inevitably occur
for a choice of mixing parameters outside of the region (\ref{ad}).
For definiteness let us refer to positive values of the asymmetry.
In this case when the total asymmetry has reached values 
$L^{(\alpha)}\gg L_t$, the resonant momentum of anti-neutrinos is 
given by the expression (\ref{yres}). After it has reached its 
minimum value $y_{\rm res}^{\rm min}$, it starts to grow again
eventually passing  all of the anti-neutrino distribution at $y>y_{\rm res}^{\rm min}$
\footnote{We recall that at this stage the resonant momentum of neutrinos is 
by far in the tail of the distribution and continues to increase further
while the asymmetry grows and therefore neutrinos are completely out of the game.}.
In the adiabatic limit all anti-neutrinos at $y=y_{\rm res}$ are converted.
It is then easy to find the following rate equation for 
the neutrino asymmetry \cite{fv2}: 
\begin{equation}
\left({dL_{\nu_{\alpha}} \over dt}\right)_{\rm ad} = 
{y^2_{\rm res}\over 4\,\zeta(3)}\,
\left[f_{\bar{\nu}_{\alpha}}(y_{\rm res})-f_{\bar{\nu}_{s}}(y_{\rm res})\right]\,
{dy_{\rm res}\over dt}
\end{equation}
In the case of negative asymmetry one has to make the replacement
$[f_{\bar{\nu}_{\alpha}}-f_{\bar{\nu}_s}]\rightarrow [f_{\nu_s}-f_{\nu_{\alpha}}]$.

In order to incorporate non adiabatic effects, which can be
important if $\sin^2 2\theta$ is small enough, we will use
the standard Landau-Zener approximation \cite{lz}. 
In this case, due to the quantum level crossing between matter eigenstates, not all 
active anti-neutrinos will be converted into sterile anti-neutrinos.
The number of active anti-neutrinos that undergo level crossing and are not
converted is given by the well known `jumping probability'
$P \equiv e^{-\pi \gamma_r/2}$ where 
$\gamma_r\equiv |2\dot{\bar{\theta}}_m\,\bar{{\ell}}_m|_{\rm res}^{-1}$
is the {\em adiabaticity parameter} at the resonance and 
$\bar{\theta}_m$ and $\bar{{\ell}}_m$
are respectively the mixing angle and the oscillation length in matter given by:
\begin{equation}
{\sin 2\bar{\theta}_m}={\sin 2\theta_0\over 
\sqrt{\sin^2 2\theta_0 +(\cos 2\theta_0 - b -a)^2}} 
\end{equation}
\begin{equation}
\bar{{\ell}}_m={2\,y\,T/ |\delta m^2|\over 
\sqrt{\sin^2 2\theta_0 +(\cos 2\theta_0 - b -a)^2}}
\end{equation}
Since we are considering a regime of high
values of the asymmetry ($L\gg L_t$), one can neglect the 
finite temperature term $b$ in the effective potential and thus one can find 
easily the following expression for the adiabaticity parameter:
\begin{equation}
\gamma_r={|\delta m^2|\over 2y_{\rm res}\,T_{\nu}}
\,{\sin^2\,2\theta_0\over |da/dt|_{\rm res}}
\end{equation}
For computational purposes this can be then transformed with easy 
algebraic arrangements making use of the Eq. (\ref{yres}), defining
$\alpha\equiv -d\ln L_{\nu_{\alpha}}/d\ln T_{\nu}>0$, replacing time derivative
with neutrino temperature derivative (using the expression (\ref{dTdt}) ) 
and finding in the end:
\begin{equation}\label{gamma}
\gamma_{\rm res}\simeq 1.8\times 10^{10}\,\sqrt{10.75\over g_{\rho}(T_{\nu})}\,\left({T_{\nu}\over T}\right)^2\,
{\sin^2 2\theta_0\,L_{\nu_{\alpha}}\,T_{\nu}\over |4-\alpha|}
\end{equation}
For $\alpha=4$ the adiabaticity parameter is infinite and this is
perfectly understandable because  when this happens, from the Eq. (\ref{yres}), one can see that 
$y_{\rm res}=y_{\rm res}^{\rm min}$ and $dy_{\rm res}/dt=0$
\footnote{Thus in this precise moment, and only in this 
moment $L\propto T^{-4}$ (exactly). After this 
moment a power-law with $\alpha=4$ cannot hold exactly, 
since if all neutrinos are adiabatically MSW 
converted, the resonant momentum has to start to grow to sustain the asymmetry growth. 
However while $L \ll 0.1$, $L \propto T^{-4}$ (approximately)
since the resonance need only move very slowly to produce
the required asymmetry.
One could measure the average value of $\alpha$ fitting the numerical
solutions if one is really interested in this quantity.
See \cite{Dolgov} for a detailed analysis on this 
(in our opinion academic) issue.}. 
The Landau-Zener approximation can be started from  
$y_{\rm res}\simeq y_{\rm res}^{\rm min}$ and in this case the rate 
equation for the asymmetry will be given by:
\begin{equation}
{dL_{\nu_{\alpha}} \over dt}=
\left({dL_{\nu_{\alpha}} \over dt}\right)_{\rm ad}\left(1 - 
e^{-{\pi\over 2}\,\gamma_{\rm res}} \right)
\end{equation}
This equation can be solved numerically and clearly is much less CPU time consumming then
the QKE's. The results from the Landau-Zener approximation for the 
final value of the asymmetry are shown in {\bf figures 5a,5b} with dashed lines
and we find a good agreement with the QKE's in both cases $\alpha=e$ and $\alpha=\mu,\tau$.
 The Landau-Zener approximation provides useful physical insight into the numerical results.
First of all we can understand equation (\ref{ad}) 
for the adiabatic region looking at the expression
(\ref{gamma}) for the adiabaticity parameter. The jumping probability 
starts to be non-negligible for $\gamma_r \lesssim 2$. The adiabaticity decreases
(on average) during neutrino asymmetry (MSW dominated) growth, since initially $\alpha=4$ 
while when the growth stops at $T=T_{\rm f}$ one has $\alpha=0$. 
The whole MSW regime will be then adiabatic
and the final value of neutrino asymmetry will be close to $0.375$ if 
$\gamma_{\rm res}(T_{\rm f})\gtrsim 2$. Using the expression (\ref{Tf}) for the 
final temperature one can immediately check that this condition reproduces
the condition (\ref{ad}) for the adiabaticity region in the space of mixing parameters. 
We can also try to estimate the equations (\ref{iso}) for the 
iso-asymmetry curves. For $\gamma_r^{\rm fin}\ll 1$ the final neutrino asymmetry
is given approximately by:
\begin{equation}\label{lnua}
L_{\nu_{\alpha}}^{\rm fin}\sim 0.375\,{\pi\over 2}\,\langle\gamma_r\rangle
\end{equation}
In this equation the quantity $\langle \gamma_r \rangle$ is the statistical average
of the adiabaticity parameter done with the Fermi-Dirac distribution with zero chemical
potential (assuming that all generation of the asymmetry occurs below the chemical
decoupling temperature). If one neglects the annihilations and assume $T_{\nu}=T$
and $g_{\rho}=10.75={\rm const}$, then this quantity can be expressed through the expression (\ref{gamma}) where the quantity $\alpha$ must be replaced with its averaged value 
$\langle \alpha \rangle$
and $\langle L_{\nu_{\alpha}} \rangle \simeq L_{\rm f}/2$. Again, from the expresssion (\ref{yres}) for the resonant momentum and using $y_{\rm res}^{\rm fin}\simeq 10$, 
obtains an expression for 
$T_{\rm f}\simeq (|\delta m^2|/{\rm eV}^2)^{1/4}\,(L_{\nu_{\alpha}}^{\rm fin})^{1/4}/3$,
and the the iso-asymmetry curves (\ref{iso}) are roughly reproduced when
$\langle\alpha\rangle \simeq 3$, that is a quite reasonable value considering that
$\alpha$ decreases from $\alpha=4$ when $y=y^{\rm min}_{\rm res}$ to $\alpha=0$ when
$y_{\rm res}\gtrsim 10$ an the asymmetry gets frozen to its final value.
Thus the Landau-Zener approximation provides the correct approach
to get a physical insight into the generation of neutrino asymmetry 
at low $-\delta m^2$ and small mixing angles.

The Landau-Zener approach is also useful to get a physical insight into the
effect of the electron-positron annihilations. From the espression (\ref{gamma})
for the adiabaticity parameter we can infact study the following quantity:
\begin{equation}
f(T/m_e)\equiv
\left[{\gamma_r (T=T_{\nu})\over \gamma_r }\right] =
\sqrt{{g_{\rho}\over 10.75}\,\left({T\over T_{\nu}}\right)^4}
\end{equation}
One can plot this function (see figure 1) and discover that it monotonically  increases from 1 
to its asymptotical value $f(0)\simeq 1.1$. Thus the answer is that  
{\em the effect of electron-positron annihilatons is to make the MSW 
conversions less adiabatic and thus to decrease the neutrino asymmetry generation.} 
This is an effect clearly negligible in the fully adiabatic region, since 
if $\gamma_r(T=T_{\nu}) \gg 1$ then also $\gamma_r \gg 1$ and the final value of the 
neutrino asymmetry does not change. While, considering the equation (\ref{lnua}), one
can see that for $|L_{\nu_{\alpha}}^{\rm fin}|\ll 1$ the annihilations decrease the final
value of $10\%$, just a correcting effect.

\section{Can the generation of neutrino asymmetry be observed ?}

\subsubsection{Big Bang Nucleosynthesis}

It remains to briefly discuss the impact of neutrino asymmetry generation on
cosmology. The most straightforward consequence is a 
modification of the predictions
of standard Big Bang Nucleosynthesis \cite{fv2,bfv,foot}. 
Standard BBN is modified by active - sterile
neutrino oscillations in two ways. One is by changing the energy
density, described by
$\Delta N_{\nu}^{\rho}$, and the second is by modifying 
the standard electron neutrino
distributions $[e^{p/T}+1]^{-1}$, described by $\Delta N_{\nu}^{f_{\nu_e}}$. 
This second change occurs both because an electron 
neutrino asymmetry can be generated and also 
because oscillations below the thermal
decoupling temperature at $\sim 1\,{\rm Mev}$ 
produce deviations from thermal 
equilibrium. In the case of $\alpha = \mu,\tau$,
clearly in the idealized 
case of two neutrino mixing 
considered in this paper, only the first effect is present in first approximation
\footnote{Actually a small effect on electron neutrino distribution, giving
a not zero $\Delta N_{\nu}^{f_{\nu_e}}$, is present also for $\alpha=\mu,\tau$
if one describes exactly the chemical decoupling. This because a few electron
neutrino-antineutrino annihilations will occur to restore $\mu$ (or $\tau$) 
number densities depleting electron neutrino number densities \cite{kimmo,foot}.}. 
However this effect is not significant unless a significant
neutrino asymmetry is generated before the
chemical decoupling temperature of about $3.5$ MeV.
This in turn requires relatively large values of $-\delta m^2
\stackrel{>}{\sim} 100\ eV^2$.
Even if all the asymmetry is produced before
the chemical decoupling (for $-\delta m^2\gg 100\,{\rm eV^2}$), 
one gets at maximum a 
value of $\Delta N^{\rho}_{\nu}\simeq 0.4$ \cite{foot}. 
In the case $\alpha=e$ the neutrino asymmetry generation
gives also a significant contribution 
to $\Delta N_{\nu}^{f_{\nu_e}}$ and in particular
if the generated asymmetry is positive then negative values 
of $\Delta N_{\nu}^{f_{\nu_e}}$
are possible \cite{fv2}. 
However, this effect is only important when a significant
neutrino asymmetry is produced before the
freezing of the neutron to proton ratio at $T \approx 0.75\, {\rm MeV}$.
This requires a $-\delta m^2 \stackrel{>}{\sim} 0.01 \ {\rm eV}^2$\cite{foot}.
This can also be seen from
figure 5a where the value of the asymmetry at $T = 0.75\,{\rm MeV}$
has been plotted for mixing angles large enough 
($\sin^2\,2\theta_0\gtrsim 10^{-8}$) that the generation is adiabatic
at $-\delta m^2/{\rm eV}^2\gtrsim 10^{-2}$. 
Also in the multi-flavour
case where a heavier $\nu_\tau$ (and/or
$\nu_\mu$) oscillates with the $\nu_s$ which generates a large
$L_{\nu_\tau}$ (and/or $L_{\nu_\mu}$)
asymmetry,
and subsequent oscillations of $\nu_\tau$ (and/or
$\nu_\mu$) with $\nu_e$ transfers
some of this asymmetry to the $\nu_e$ sector
(thereby generating $L_{\nu_e}$) one similarly obtains 
that significant $L_{\nu_e}$
production requires a $-\delta m^2 \stackrel{>}{\sim} 0.5 \ eV^2$
\cite{fv2,bfv,foot}.
{\em Thus for $|\delta m^2| \ll 10^{-2}\, {\rm eV}^2$, 
the asymmetry generation effects are unlikely to produce observable
deviations on the standard BBN predictions.} 

\subsubsection{Cosmic Microwave Background}

It has been pointed out how a neutrino asymmetry 
generated above the chemical decoupling
temperature would imply a $\Delta N_{\nu}^{\rho}$ 
able to change the matter-radiation
equivalence and thus modifying the height and 
position of CMBR acoustic peaks \cite{sarkar}.
However in this case such an asymmetry cannot be generated from 
active-sterile neutrino oscillations because, as already pointed out 
discussing BBN, they  generate the asymmetry 
generally below the chemical decoupling and thus produce a value of $\Delta N_{\nu}^{\rho}$ 
not high enough to produce remarkable effects. 
However a $\Delta N_{\nu}^{\rho}$ is also 
produced for mixing angles large enough (see Eq. (\ref{upper})) 
that a significant sterile neutrino production occurs. 
This also modifies the matter-radiation equivalence time and
gives effects on CMBR anisotropies \cite{raffelt}. 
In the case of negative $-\delta m^2$
the neutrino asymmetry generation can 
modify the sterile neutrino production, especially
in multiflavor mixing. Thus this indirect 
effect of neutrino asymmetry generation has 
effects on CMBR. 
In other words a calculation of $\Delta N^{\rho}_{\nu}$ 
including the account of asymmetry
generation is needed to derive the correct effects on CMBR anisotropies.

\subsubsection{Relic Neutrinos}

We have seen that the effects of neutrino asymmetry generation on BBN
for low values of $-\delta m^2$ are not significant, while on CMBR they
can be only indirect. However it is still remarkable 
that neutrino asymmetry 
generation would lead to a relic neutrino background in 
which mainly anti-neutrinos
or neutrinos are present. At the moment nobody
has been able to experimentally detect 
relic neutrinos.  However it has been 
noted in \cite{smirnov} that
the presence of an asymmetry in the relic neutrino background can 
have observable effects in future neutrino detectors. If very high energy
$\alpha$-neutrinos are produced 
from cosmological sources at high redshift, their 
vacuum probability oscillation can be changed by 
matter effects due to the presence of an asymmetry
in the relic background. Thus, if one 
knows the vacuum probability and the fluxes and 
energy spectrum of produced neutrinos, it is possible in principle 
to observe
a change in the oscillation probability induced by matter effects. 
These effects
depend crucially on the value of $\eta^{(\alpha)}_0\equiv 
h_0\,L^{(\alpha)}$, where 
$h_0=4/11$ is the dilution factor at the present. 
Thus the maximum value of $\eta^{(\alpha)}_0$
that can be generated
from active-sterile neutrino oscillations is $\simeq 0.27$. 
According to the analysis 
presented in \cite{smirnov}, such a value can give 
at maximum effects of about $0.5\%$ 
in the change of vacuum probability which is
probably unobservably small.
Nevertheless,
it shows how the possibilities of detecting 
relic neutrino background can be changed by the presence of an 
asymmmetry produced by active-sterile neutrino oscillations and thus 
the issue of relic neutrino background detection should be 
reconsidered in light of this and deserves further investigations. 

\section{Conclusions}

We have shown that the generation of large
neutrino asymmetry via active - sterile
neutrino oscillations occurs for much lower values
of $|\delta m^2|$ than previously supposed.
We have made a detailed study of the
final asymmetry values in the $\sin^2 2\theta_0\leq 10^{-7}$ parameter region.
The evolution of neutrino asymmetry in this parameter region
can be understood quite simply by means of the
Landau-Zener approximation that made possible to derive analytical
results in good agreement with the numerical ones. 
For very small $-\delta m^2 \ll 0.01 \ eV^2$,
the oscillations generate the asymmetry too late
to have significant effects on BBN. The account of neutrino 
asymmetry generation can modify the sterile neutrino production 
(this is especially true when considered within a multiflavor mixing scheme)
and thus has to be taken into account when studying the effects on CMBR anisotropies.
Properties of relic neutrino background are also modified by neutrino asymmetry generation
and this observation should certainly encourage new investigations on relic neutrinos 
detection.

\vskip 0.5cm
\noindent
{\bf Acknowledgements}
For the numerical calculations we used the 
machines of the MARC centre at the University of Melbourne. 
We thank the staff members for technical support. One of us, PDB,
wishes to thank A.D. Dolgov for nice discussions during the
XXXVIth Rencontres de Moriond session on Electroweak 
Interactions and Unified Theories held in Les Arcs.

\newpage
{\bf Figure Captions}

\vskip 0.5cm
\noindent
{\bf Figure 1.} Effect of electron-positron annihilations on the degrees of freedom,
on neutrino temperature and on the adiabaticity parameter.  
\vskip 0.5cm
\noindent
{\bf Figure 2.} Evolution of $L^{(e)}=2\,L_{\nu_e}+\tilde{L}$ 
with (photon) temperature for $\sin^2\,2\theta_0=10^{-8}$
and values of $-\delta m^2$ as indicated. The solid lines
are the QKE's solutions while the dotted lines employ
an adiabatic description for the growth in the MSW dominated regime.
The initial value is $L^{(e)}=\tilde{L}=5\times 10^{-10}$.

\vskip 0.5cm
\noindent
{\bf Figure 3.} The same as in figure 2 but for $\sin^2\,2\theta_0=10^{-7}$
and a different choice of values for $-\delta m^2$. 
For $-\delta m^2/{\rm eV}^2=10^{-8}$ we also show an example for a different choice 
of the initial asymmetry.
\vskip 0.5cm
\noindent
{\bf Figure 4.} An example of evolution with temperature for $L^{(\mu)}$.
The mixing parameters are $\sin^2\,2\theta_0=-\delta m^2/{\rm eV}^2=10^{-7}$.

\vskip 0.5cm
\noindent
{\bf Figure 5.}
Final values of the neutrino asymmetry in the case $\alpha=e$ (a) and
$\alpha=\mu$ (b). Solid lines are solutions from the QKE's. Dashed lines
are the solution with Landau-Zener approximation. Dotted lines are the 
adiabatic limit. In the figure 5a we also show with a thin solid line
the value of neutrino asymmetry at $T=0.75\,{\rm MeV}$ (for mixing angles
$\gtrsim 10^{-8}$) that is approximately the freezing 
temperature of the neutron to proton ratio. 
\vskip 0.5cm
\noindent
{\bf Figure 6.}
Adiabatic region and iso-$|L_{\nu_{e}}^{\rm fin}|$ curves.
\vskip 0.5cm
\noindent
{\bf Figure 7.} The same as in figure 2 but for $\sin^2\,2\theta_0=4\times 10^{-10}$
and a different choice of values for $-\delta m^2$. 
For $-\delta m^2=10^{-3}\,{\rm eV}^2$ the MSW dominated regime never starts.

\newpage
\epsfig{file=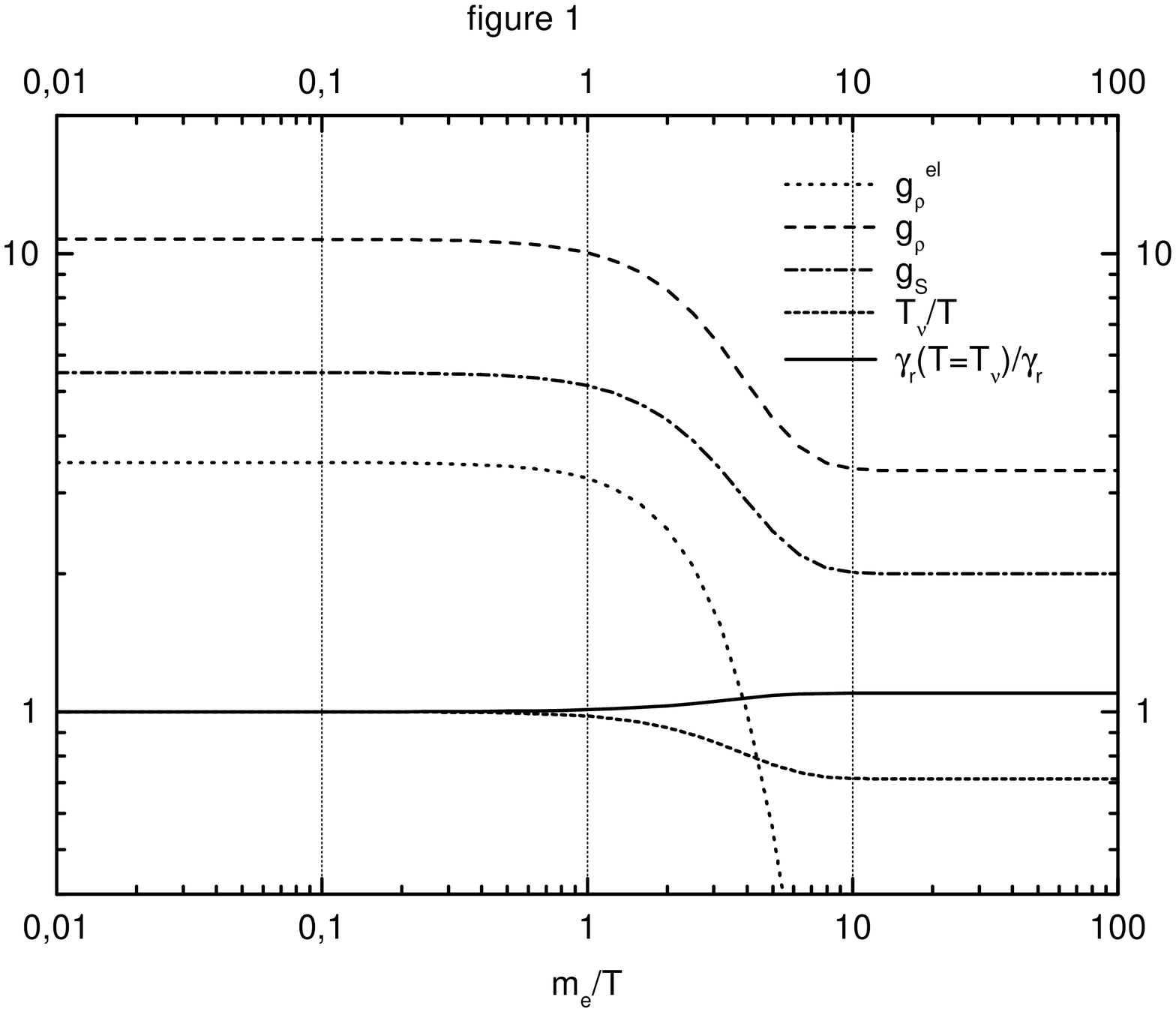,width=15cm}
\newpage
\epsfig{file=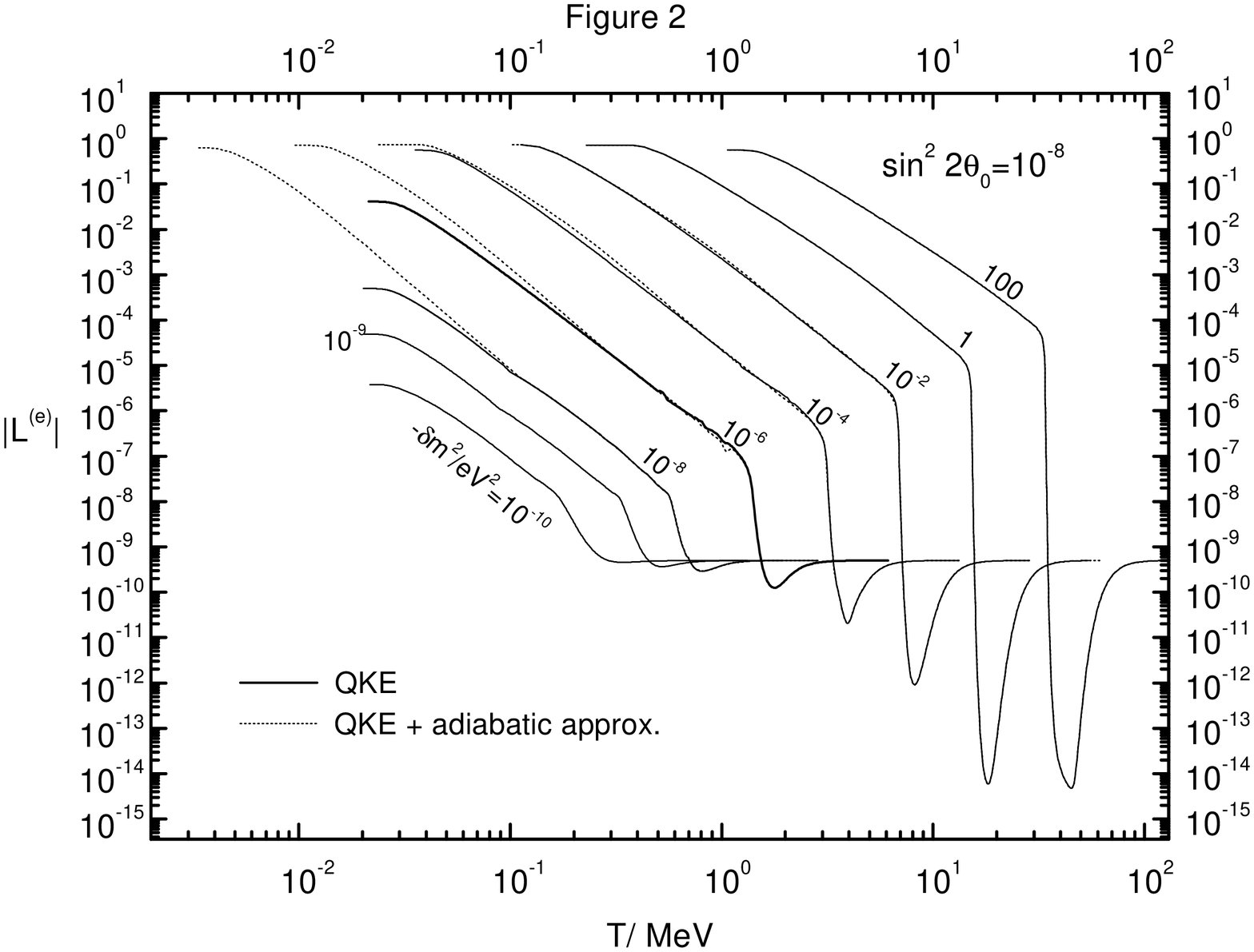,width=15cm}
\newpage
\epsfig{file=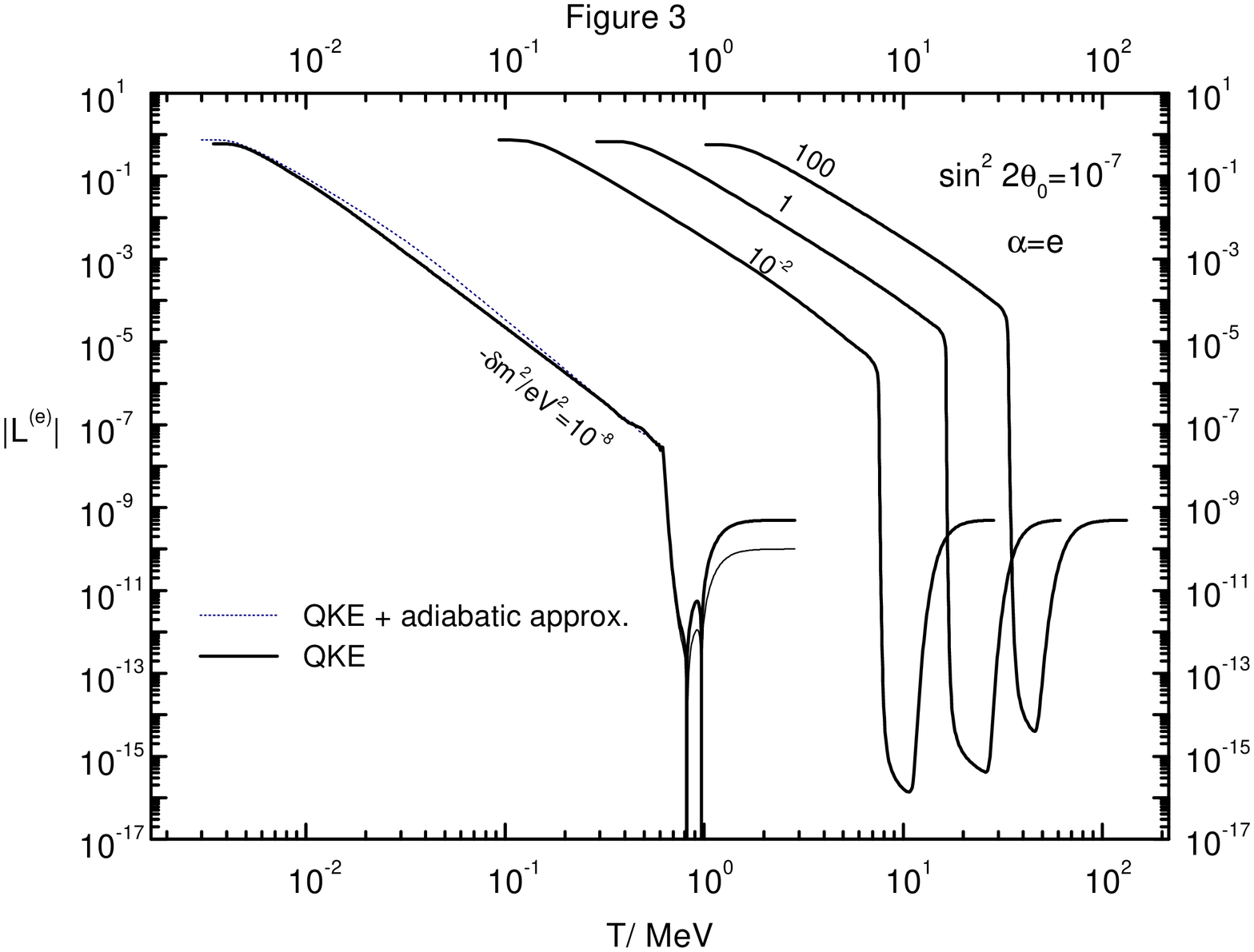,width=15cm}
\newpage
\epsfig{file=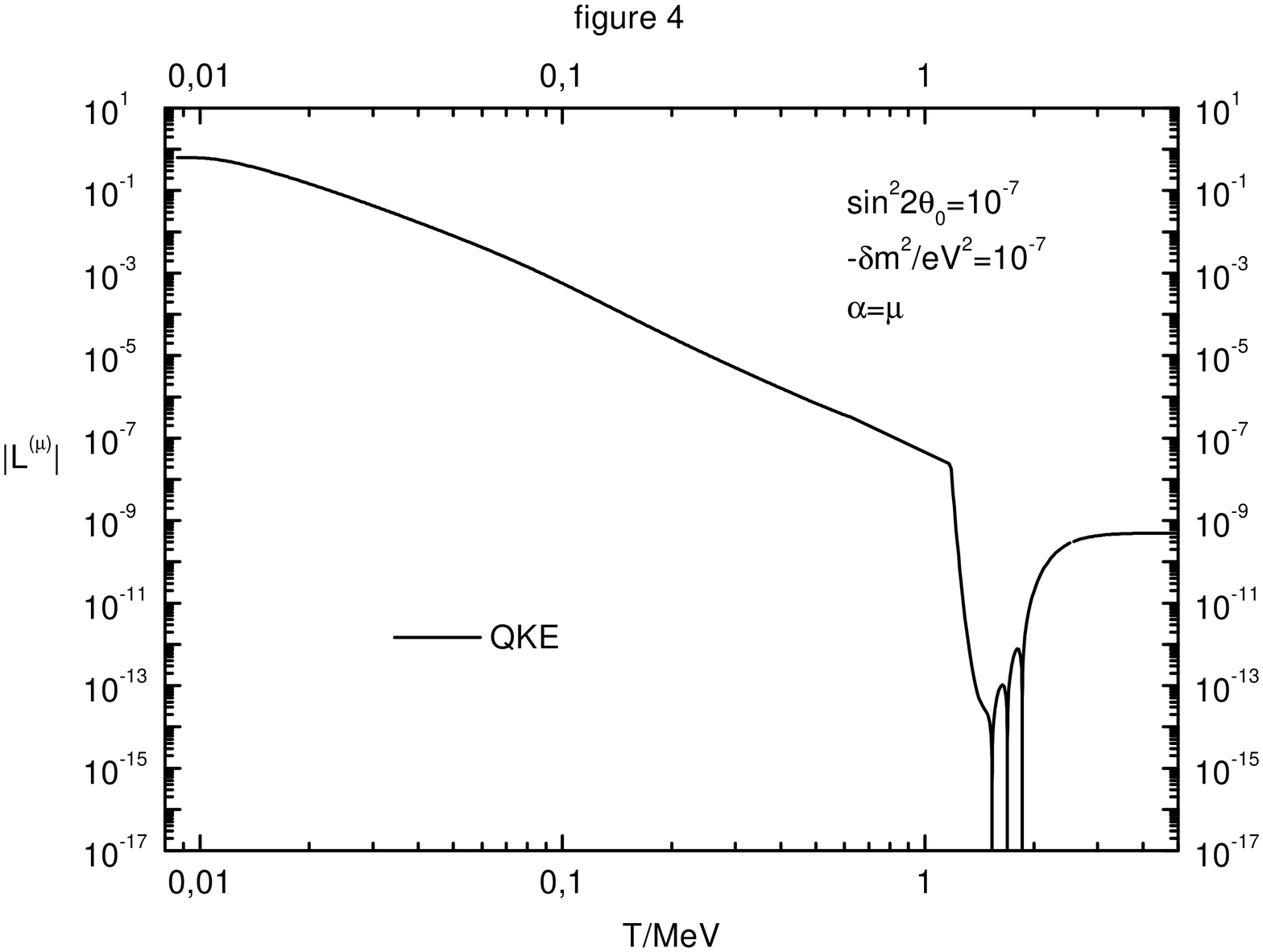,width=15cm}
\newpage
\epsfig{file=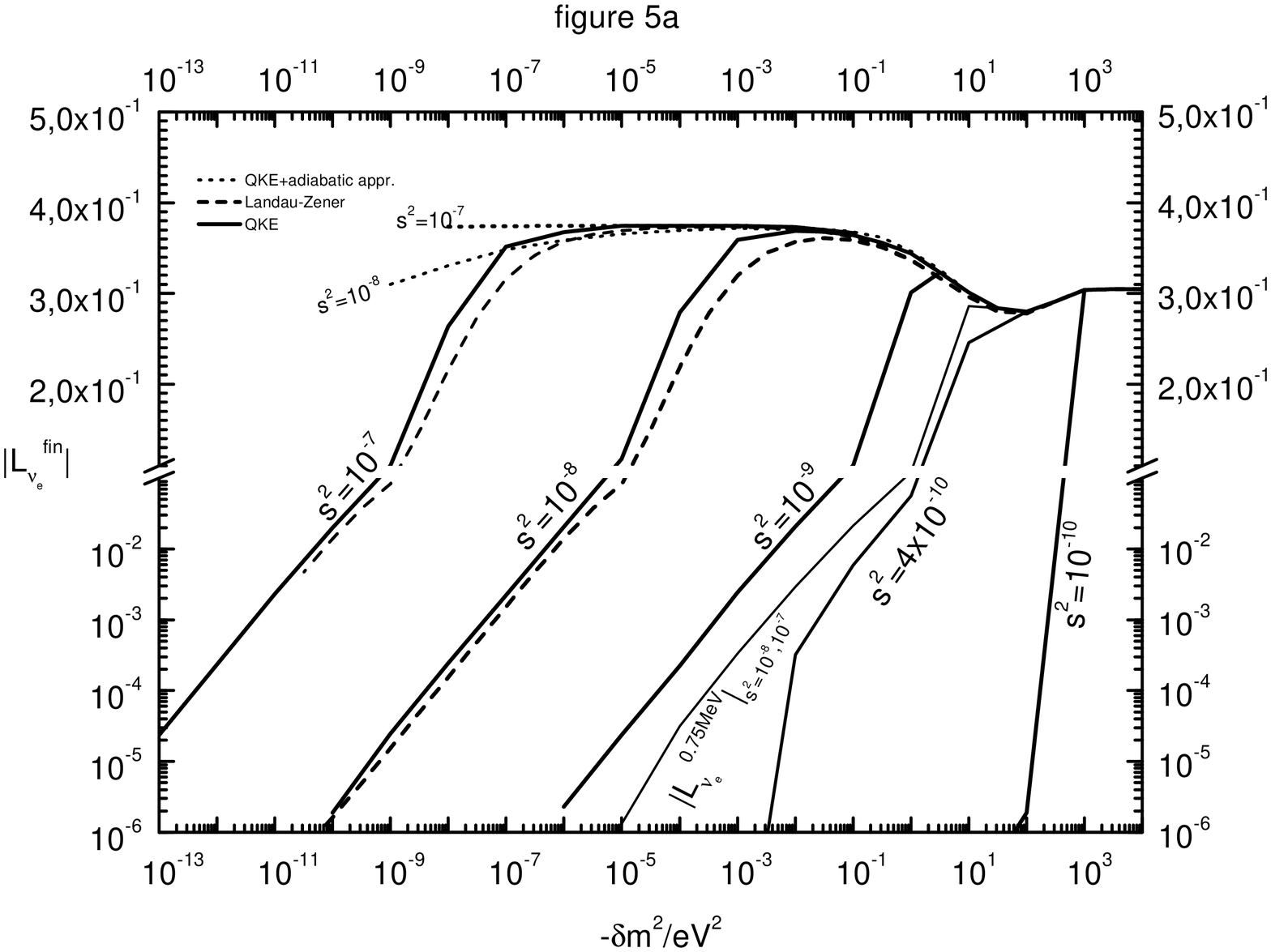,width=15cm}
\newpage
\epsfig{file=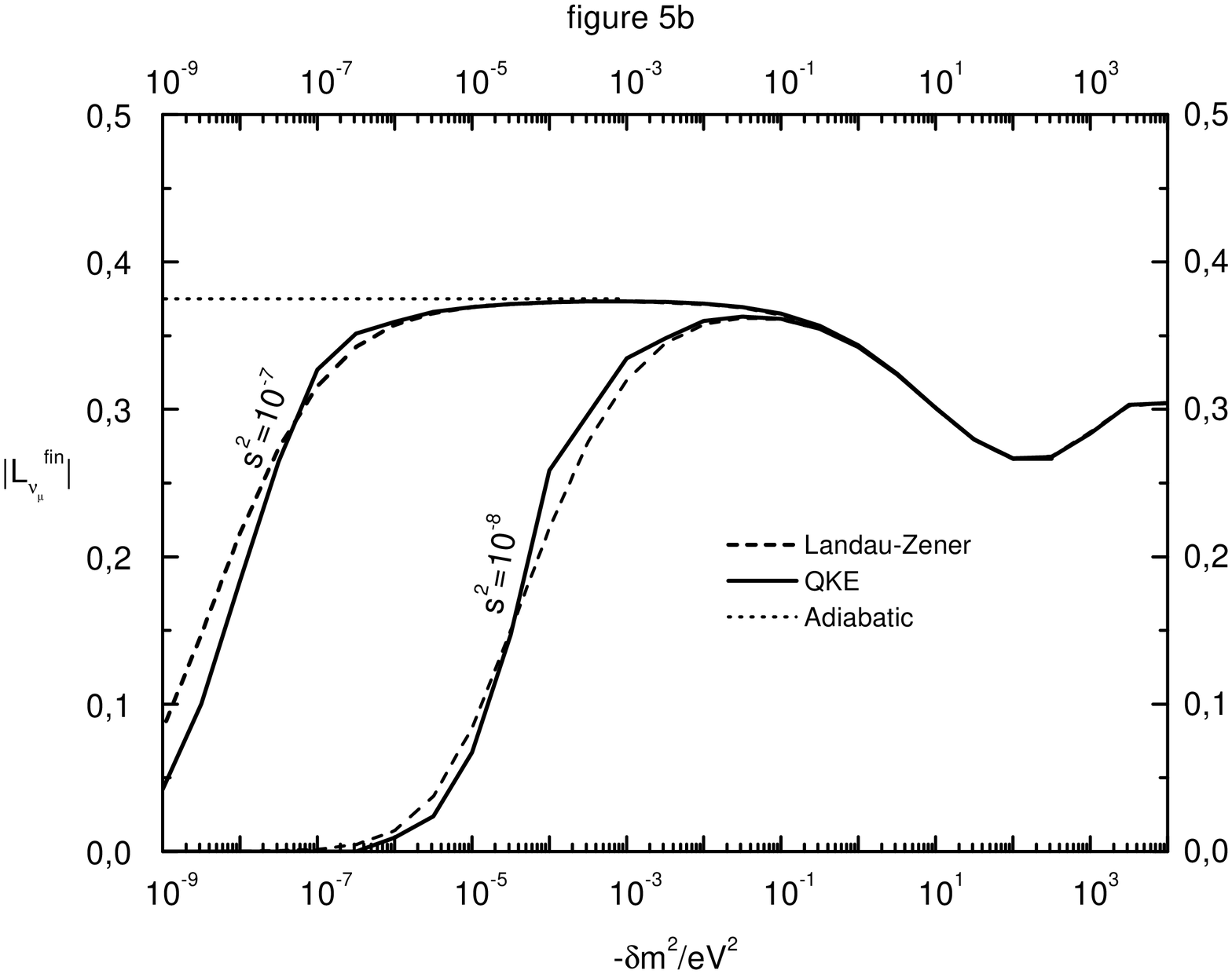,width=15cm}
\newpage
\epsfig{file=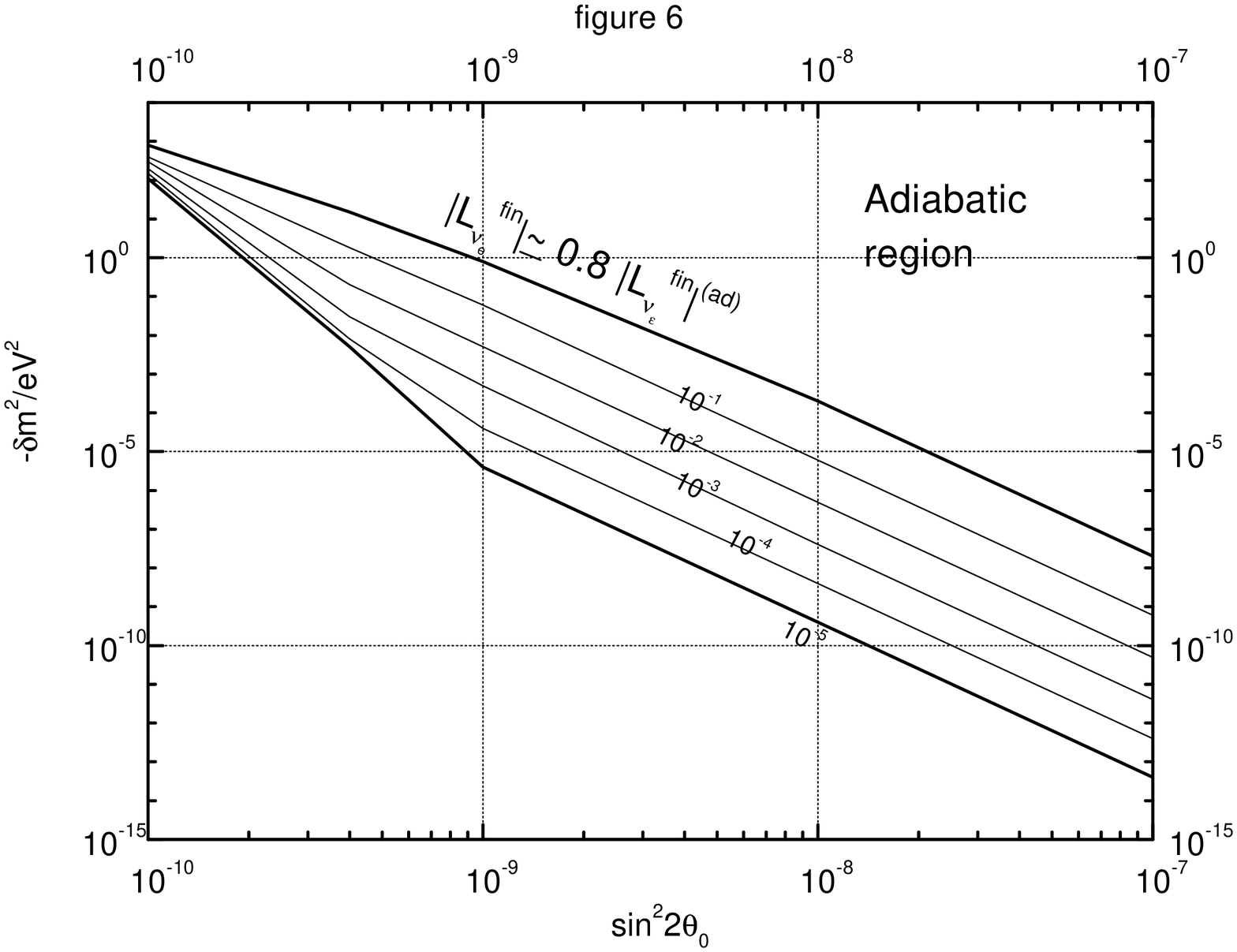,width=15cm}
\newpage
\epsfig{file=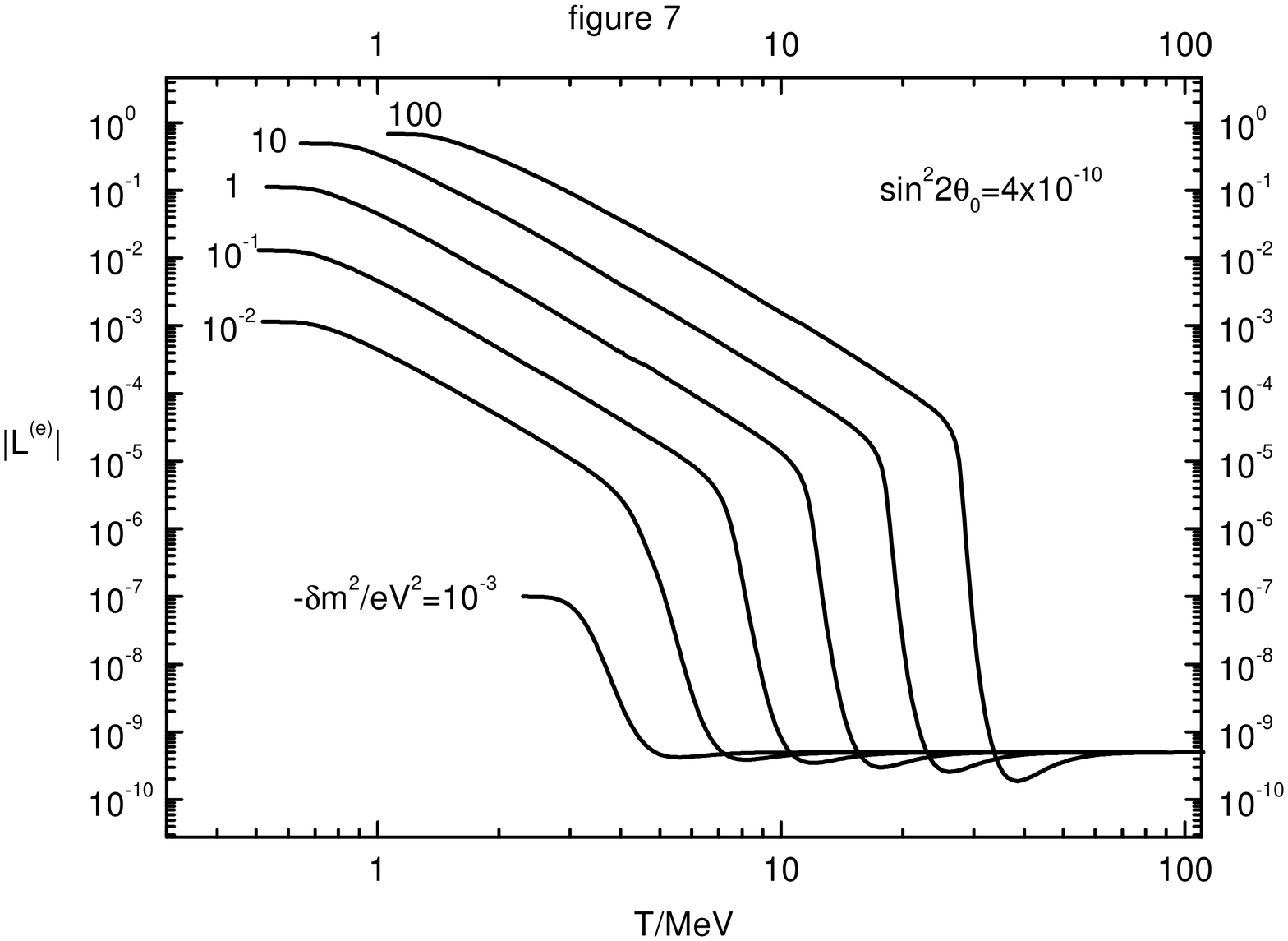,width=15cm}


\begin{thebibliography}{10}

\bibitem{ftv}
R.~Foot, M.~J.~Thomson and R.~R.~Volkas,
Phys.\ Rev.\ {\bf D 53} (1996) 5349
[hep-ph/9509327].

\bibitem{fv1}
R.~Foot and R.~R.~Volkas,
Phys.\ Rev.\ {\bf D 55}, 5147 (1997)
[hep-ph/9610229].

\bibitem{fv2}
R.~Foot and R.~R.~Volkas,
Phys.\ Rev.\ {\bf D 56}, 6653 (1997)
[hep-ph/9706242].

\bibitem{f98}
R.~Foot, Astropart.\ Phys.\ {\bf 10} (1999) 253 [hep-ph/9809315].

\bibitem{fv}
R.~Foot and R.~R.~Volkas,
Phys.\ Rev.\ Lett.\ {\bf 75} (1995) 4350
[hep-ph/9508275].

\bibitem{dll}
P.~Di Bari, P.~Lipari and M.~Lusignoli,
Int.\ J.\ Mod.\ Phys.\ {\bf A15}, 2289 (2000)
[hep-ph/9907548].

\bibitem{shi}
X.~Shi,
Phys.\ Rev.\ {\bf D 54}, 2753 (1996)
[astro-ph/9602135].

\bibitem{bfv}
N.~F.~Bell, R.~Foot and R.~R.~Volkas,
Phys.\ Rev.\ {\bf D 58}, 105010 (1998)
[hep-ph/9805259].

\bibitem{foot}
R.~Foot,
Phys.\ Rev.\ {\bf D 61}, 023516 (2000)
[hep-ph/9906311].

\bibitem{domains}
P.~Di Bari,
Phys.\ Lett.\ {\bf B482}, 150 (2000)
[hep-ph/9911214].

\bibitem{ropa2}
P.~Di Bari and R.~Foot,
Phys.\ Rev.\ {\bf D 63}, 043008 (2001) [hep-ph/0008258].

\bibitem{ropa}
P.~Di Bari and R.~Foot, Phys.\ Rev.\ {\bf D 61} (2000) 105012 [hep-ph/9912215].
See also
K.~Enqvist, K.~Kainulainen and A.~Sorri, Phys.\ Lett.\ {\bf B464} (1999) 199 [hep-ph/9906452],
for the case of the mean momentum toy model. 

\bibitem{ben}
B. Mc Millan, honours thesis (2000), School of Physics, 
University of Melbourne.

\bibitem{BoomMax} 
A.H. Jaffe et al., astro-ph/0007333;
S.~Esposito, G.~Mangano, A.~Melchiorri, G.~Miele and O.~Pisanti,
Phys.\ Rev.\ D {\bf 63} (2001) 043004
[astro-ph/0007419];
J.~P.~Kneller, R.~J.~Scherrer, G.~Steigman and T.~P.~Walker, astro-ph/0101386.


\bibitem{QKE}
A.~D.~Dolgov, Yad.\ Fiz.\ {\bf 33} (1981) 1309;
Sov. J. Nucl. Phys. {\bf 33}, 700 (1981);
R.~A.~Harris and L.~Stodolsky, Phys.\ Lett.\ {\bf B116} (1982) 464;
L.~Stodolsky, Phys.\ Rev.\ {\bf D 36}, 2273 (1987);
B.~H.~McKellar and M.~J.~Thomson, Phys.\ Rev.\ {\bf D 49}, 2710 (1994).




\bibitem{comment}
P.~Di Bari, R.~Foot, R.~R.~Volkas and Y.~Y.~Wong,
hep-ph/0008245.

\bibitem{bvw}
N.~F.~Bell, R.~R.~Volkas and Y.~Y.~Wong,
Phys.\ Rev.\ {\bf D 59} (1999) 113001
[hep-ph/9809363].

\bibitem{bar}
R.~Barbieri and A.~D.~Dolgov,
Phys.\ Lett.\ {\bf B237} (1990) 440.

\bibitem{ekm}
K.~Enqvist, K.~Kainulainen and J.~Maalampi, Nucl.\ Phys.\ {\bf B349} (1991) 754.

\bibitem{smirnov} 
C.~Lunardini and A.~Y.~Smirnov, hep-ph/0012056.

\bibitem{Notzold}
D.~Notzold and G.~Raffelt, Nucl.\ Phys.\ {\bf B307} (1988) 924.



\bibitem{lz}
L.D. Landau, Phys. Z. Sowjetunion {\bf 2}, 46 (1932);
C.~Zener, Proc.\ Roy.\ Soc.\ Lond.\ A {\bf 137} (1932) 696;
which has been applied to neutrino oscillations in
matter by:
W.~C.~Haxton, Phys.\ Rev.\ Lett.\ {\bf 57} (1986) 1271;
S.~J.~Parke, Phys.\ Rev.\ Lett.\ {\bf 57} (1986) 1275;
A.~Dar, A.~Mann, Y.~Melina and D.~Zajfman, Phys.\ Rev.\ D {\bf 35} (1987) 3607.



\bibitem{Dolgov}
R.~Buras and D.~V.~Semikoz, hep-ph/0009266;
A.~D.~Dolgov, hep-ph/0102125.

\bibitem{kimmo}
K.~Enqvist, K.~Kainulainen and M.J.~Thompson, Nucl.\ Phys.\ {\bf B373}, (1992) 498.
\bibitem{sarkar}
J.A.~Adams and S.~Sarkar, preprint OUTP-98-70P and talk presented at
the workshop on {\sl The Physics of Relic Neutrinos}, Trieste,
September 1998; 
J.~Lesgourgues and S.~Pastor, Phys.\ Rev.\ D {\bf 60} (1999) 103521 [hep-ph/9904411];
W.~H.~Kinney and A.~Riotto, Phys.\ Rev.\ Lett.\ {\bf 83} (1999) 3366 [hep-ph/9903459].

\bibitem{raffelt} S.~Hannestad and G.~Raffelt, Phys.\ Rev.\ D {\bf 59} (1999) 043001
[astro-ph/9805223].

\end{thebibliography}
\end{document}